\documentclass[10pt,aps,cite,citesort,fullpage]{revtex4}

\usepackage{setspace}
\usepackage{apjfonts}
\usepackage{epsfig,graphicx}
\usepackage{citesort}
\usepackage{pifont}
\usepackage{amssymb}
\usepackage{amsmath}
\usepackage{latexsym}
\usepackage{verbatim}
\usepackage{array}
\usepackage{xspace}
\usepackage{colordvi}
\usepackage{wasysym}
\usepackage{marvosym}

\newcommand {\be}{\begin{equation}} 
\newcommand {\ee}{\end{equation}}

\def \e{\epsilon}

\def \be{\begin{equation}}
\def \ee{\end{equation}}
\def \bea{\begin{eqnarray}}
\def \eea{\end{eqnarray}}
\def \zn{z_{\mbox{\tiny N}}}
\def \zb{z_{\mbox{\tiny B}}}

\def \kboltz{k_{\mbox{\tiny B}}}

\def \Del{\Delta}

\def \e{\mbox{e}}
\def \En{E_{\mbox{\tiny N}}}

\def \enn{\eps_{\mbox{\tiny NN}}}

\def \D{\partial}

\def \En{E_{\mbox{\tiny N}}}

\def \lbar{\overline{\ell}}
\def \Tg{T_{\mbox{\tiny G}}}
\def \Tf{T_{\mbox{\tiny F}}}

\def \s{\sigma}

\def \eps{\epsilon}

\def \d{\delta}

\def \kF{k_{\mbox{\tiny F}}}

\def \Amax{A_{\mbox{\tiny MAX}}}
\def \Tfo{T_{\mbox{\tiny F}}^o}
\def \Tfosq{T_{\mbox{\tiny F}}^{o\: 2}}

\def \Ast{A^{\ast}}
\def \etast{\eta^{\ast}}
\def \etabar{\overline{\eta}}
\def \l{\lambda}
\def \lx{\lambda_x}
\def \ly{\lambda_y}
\def \lz{\lambda_z}
\def \pbar{\overline{p}}

\def \nL{n_{\mbox{\tiny L}}}
\def \Nnuc{N_{\mbox{\tiny NUC}}}
\def \bA{b_{\mbox{\tiny A}}}

\newlength{\saveparindent}
\newcommand {\eg   }{\textit{e.g.}\xspace}%
\newcommand {\ie   }{\textit{i.e.}\xspace}

\newcounter{saveeqn}%
\newcommand{\alpheqn}{\setcounter{saveeqn}{\value{equation}}%
\stepcounter{saveeqn}\setcounter{equation}{0}%
\renewcommand{\theequation}{\mbox{\arabic{saveeqn}\alph{equation}}}}%
\newcommand{\reseteqn}{\setcounter{equation}{\value{saveeqn}}%
\renewcommand{\theequation}{\arabic{equation}}}%

\begin{document}

\pagenumbering{arabic}
\setcounter{page}{1}

\title{The effects of non-native interactions on protein folding
  rates: Theory and simulation}

\author{Cecilia Clementi$^1$\footnote{e-mail: cecilia@rice.edu, fax: +1-713-348-3485}
and Steven  S. Plotkin$^2$\footnote{e-mail: steve@physics.ubc.ca, fax: +1-604-822-5324}
}

\affiliation{ $^1$ Department of Chemistry,
and W.~M.~Keck Center for Computational and Structural Biology,
Rice University, 6100 Main Street, Houston,
TX 77005, USA\\
$^1$Structural and Computational Biology and Molecular Biophysics,
Baylor College of Medicine, One Baylor Plaza, Houston, TX 77030, USA\\
$^2$ Department of Physics and Astronomy, University of British
Columbia, 6224 Agricultural Road, Vancouver, BC V6T1Z1, Canada}

\baselineskip=0.2cm

\date{\today }

\begin{abstract}
Proteins are minimally frustrated polymers. However, for realistic
protein models non-native interactions must be taken into account. In this
paper we analyze the effect of non-native interactions on the folding rate
and on the folding free energy barrier.
We present an analytic theory to account for the modification on the free
energy landscape upon introduction of non-native contacts, added as a
perturbation to the strong native interactions driving folding.  Our theory
predicts a rate-enhancement regime at fixed temperature, under the
introduction of weak, non-native interactions. 
We have thoroughly tested this theoretical prediction with simulations of a
coarse-grained protein model, by employing an off-lattice $C_\alpha$
model of the src-SH3 domain.
The strong agreement between results from simulations and theory confirm
the non trivial result that a relatively small amount of non-native
interaction energy can actually assist the folding to the native structure.
\end{abstract}
\vspace{5cm}

\maketitle
\normalsize

\section{Introduction}

The mechanism of protein folding is of central importance to
structural and functional biology (see
\eg~\cite{WinklerACR1998,FershtBook,CreightonBook,PlotkinSS02:quartrev1,PlotkinSS02:quartrev2}).
An understanding of the fundamental
physical-chemical factors regulating the folding process may help
provide answers to some of the long outstanding problems in both
functional genomics and biotechnology: rational design of drugs and
enzymes, potential control of genetic diseases, and a deeper
understanding of the connection between biological structure and
function are among the applications that may benefit from advances in
protein folding.

Theoretical and computational studies have recently achieved noticeable success
in reproducing various features of the folding mechanisms of several
small to medium-sized fast-folding proteins (see \eg~\cite{KaranicolasPNAS2003,Shea2001,sorenson_head-gordon:00:jcb,ShakhnovichPNAS2002,shoemaker_wang_wolynes:99:jmb,Clementi2003:JMB,KayaH03,sorenson_head-gordon:02:jcb}); at the same time, the improved
spatial and temporal resolution of recent experimental techniques is now allowing researchers to combine
theoretical and experimental data to give a more robust characterization of the folding
free energy landscape~\cite{GruebeleJBP2002,lapidus_eaton_hofrichter:00:pnas,schuler_lipman_eaton:02:nature,PandePNAS2003,PandeGruebeleNature2002,EatonJMB2003}.
However in spite of these recent successes, a microscopically detailed observation of the individual
conformational motions that occur during folding remains elusive.
A knowledge of the time-dependence of every degree of freedom in the system is,
however, not of inherent interest, since no additional insight to the underlying
physics of the folding process is gained from this information by itself.
Nor is any particular degree of freedom especially important to folding, because
the transition involves the cooperation of many weakly (non-covalently) interacting
constituents. For these reasons a statistical description of the process of folding,
in terms of the behavior of an ensemble of systems, is appropriate for distinguishing
general (self-averaging) properties from sequence-specific ones~\cite{BryngelsonJD95}.
The characterization of the folding process in statistical mechanical terms can
pinpoint crucial questions that may be computationally or experimentally addressed in more detail.

The idea of considering ensemble properties to characterize the folding landscape
underpinned studies of the transition state and folding mechanism as arising from the native
state topology~\cite{ShoemakerBA97,AlmE99,MunozV99,FinkelsteinAV99:pnas,Clementi2000:PNAS,Clementi2000:JMB,SheaJE00,Clementi2001:JMB,Clementi2003:JMB}.
As a general rule, the transition state structure does not differ
dramatically between homologous proteins~\cite{PlaxcoKW00:jmb,Gierash2001:COSB,Baker2000:Nature}, and any
exceptions are fairly readily explained~\cite{FergusonN99,KimDE00}.
Consistent with the above-mentioned notions of self-averaging, folding
rates of homologous proteins are seldom seen to differ by more
than an order of magnitude when tuned to the same
stability~\cite{Mines96,PlaxcoKW00:biochem}. This indicates that the
folding free energy barrier is not particularly sensitive to the
details of sequences folding to a given native structure, but
depends rather on more general features of that ensemble of sequences,
including the kinetic accessibility of that native structure.
In this sense, the topology of the native structure
largely determines the folding free energy barrier for those homologous
sequences~\cite{PlaxcoKW00:biochem}.

These ideas motivated many studies of folding rates and mechanisms
using so-called G\={o} models~\cite{GoN75}, which neglect interactions
not present in the native state. In these studies the possibility of structure
prediction is traded for the possibility of rate and mechanism
prediction. Moreover, because of the robustness of rate and mechanism
for homologous proteins, the coarse-graining of the G\=o model (\ie removing the
molecular details of side-chains and solvent) is often assumed a reasonable
approximation.


Topology-based approaches seek to predict mechanism by calculating
$\phi$-values~\cite{fersht1986:nature,MatouschekA90} or analogous
quantities,  which in an accurate theory give values that correlate
with experiment for the measured cases. Occasionally one finds
residues whose $\phi$-values are negative. This is most likely due to
the presence of non-native contacts that stabilize the transition
state, but cannot be present in the native state. The presence of
non-native interactions in the transition state is supported by
all-atom simulations using a Charmm-based effective energy function,
where it was found that about $20-25\%$ of the energy in the
transition state arose from non-native contacts~\cite{PaciE02bj}.

Hence for a more realistic protein potential energy function,
non-native interactions
must be taken into account. In this paper we analyze the effect of
increasing the strength of non-native interactions on the folding rate
as well as the free energy barrier.
Non-native interactions are introduced as additional
contacts between pairs of residues not in contact in the
native structure, which are allowed to have a non-zero mean and a
non-zero variance. The non-native interactions are
added perturbatively to the G\={o} model: all non-native contacts are given
a random energy with mean $\enn$ and a variance $b^2$ which is progressively
increased to examine more frustrated proteins, while the native
contact energies are all held fixed to the same
number. The limiting case of $\enn=0$ and $b=0$ corresponds to the
plain G\=o model.
This procedure essentially preserves the stability of the native state, where
approximately no non-native interactions are present.
However, the stability of the unfolded state is lowered (as shown
in \S\ref{sec:rate_theory} and \S\ref{sec:thermo} of this paper).

At first glance one would expect that introducing progressively larger
non-native contact energies to an otherwise energetically unfrustrated
G\=o protein would slow the folding rate, for straightforward
reasons: It would seem that ``noise'' in the system would make the
native basin harder to recognize. One might argue by analogy that it
is easier to read a page of text without random misspellings.
However, the folding rate has been predicted to initially
increase under the introduction of weak, non-native interactions,
added as a perturbation to the strong native interactions driving
folding~\cite{PlotkinSS01:prot}. This was a fold-independent result
derived from general principles of energy landscape theory.
This prediction was subsequently verified in simulations of a
$36$-mer lattice model~\cite{FanK02}, as well as off-lattice molecular
dynamics simulations of Crambin, in which attractive non-native
contacts were successively added~\cite{CieplakM02}. Independently,
it was found that non-native interactions were present
in the transition state of a 28-mer lattice-model protein with
side-chains, and increased the
folding rate when strengthened~\cite{LiL00}. Similar observations were also
seen in 2-dimensional 24-mer lattice models~\cite{TreptowWL02}.
A different computational study on a 36-mer lattice-model protein
found that at the temperature of fastest folding in simulation models,
the folding rate monotonically decreases with increasing ruggedness~\cite{FanK02}
(the temperature of fastest folding of course varies with the ruggedness). However
this typically barrierless regime is rarely seen in the
laboratory~\cite{SabelkoJ99,GruebeleM99}.

The prediction that strengthening  non-native interactions that were
initially weak would accelerate folding is also consistent with
experimental observations that strengthening non-specific hydrophobic
stabilization in $\alpha$-spectrin Src homology 3 (SH3) domain sped up folding
(and unfolding) for that protein~\cite{VigueraAR02}.
This result was significantly non-trivial, to the extent that the
experimental observation was originally interpreted (mistakenly) as evidence
against the energy landscape theory.

In this paper, we test this prediction with simulations of a
coarse-grained protein model, by employing an off-lattice $\mbox{C}_\alpha$
model (see \eg~\cite{HoneycuttJD92,Clementi2000:JMB}) of 
the SH3 domain of {\it src tyrosine-protein kinase} (src SH3).
domain. We use a Hamiltonian
function that has tunable amounts of non-native energy (see
Appendix~\ref{sect:app-sim} for details).  The results from simulations are
compared with the predictions of an improved
version of the existing theory~\cite{PlotkinSS01:prot}. The theory is
improved by introducing a finite-size treatment of packing fraction as a function of
polymer length, which takes better account of the polymer physics
involved in collapse as folding progresses.
Moreover, the previous study treated the rate enhancement at fixed
stability. Here we show a perhaps even less intuitive result, namely
that the rate-enhancement can happen at fixed temperature, and we derive
the conditions required for this to happen.

As the strength of non-native interactions is increased to larger
values, we find that eventually the folding rate decreases drastically, as expected.
In the limit of large non-native contact energies, the chain behaves like
a random heteropolymer, having misfolded structures more stable than
the native state.

The folding mechanism is also non-trivially effected by the introduction
of non-native interactions. In this regard, the analysis of the robustness
of the folding mechanism against an increasingly strong perturbation on the
non-native interactions  can provide a critical assessment on the validity of
unfrustrated protein models for the prediction of folding mechanism, for
different protein topologies.
This analysis goes beyond the scope of the present paper and it will be
addressed separately~\cite{Plot-Clem03}.

The paper is organized as follows. In the next section (\S\ref{sec:theory})
we present the theory. After presenting the general ideas and overall
strategy (\S\ref{sec:theoryoverview}), we discuss in detail how an explicit
expression for the conformational entropy can be obtained in terms of the
packing fraction (\S\ref{sec:entropy}).
We use this result to show how thermodynamic free energy barrier is lowered
by the presence of non-native interactions (\S\ref{sec:rate_theory}).
In section \ref{sec:simulations} we test the theoretical predictions with
direct simulation of the src-SH3 domain.
We first compare the definition of reaction coordinates and the relative
approximations of theory and simulations (\S\ref{sec:Asim}); thermodynamic
(\S\ref{sec:thermo} and \S\ref{sec:temp}) and kinetic quantities
(\S\ref{sec:rate}) obtained from simulations are then quantitatively
compared with the corresponding theoretical predictions.

The strong
agreement between results from simulations and theory confirm the non
trivial result that a relatively small amount of non-native interaction
energy can actually assist the folding to the native structure.

\section{Theory of folding with non-native interactions}
\label{sec:theory}

\subsection{Definition of the general strategy}

\label{sec:theoryoverview}

Thermodynamic quantities relevant to folding may be obtained from
an analysis of the density of states in the presence of energetic
correlations~\cite{PlotkinSS02:quartrev1,PlotkinSS02:quartrev2}. In
this context we introduce two order parameters. We let $Q$ be the
fraction of contacts shared between an arbitrary structure and the
native structure, and we let $A$ be the fraction of possible
non-native contacts present in that structure, i.e. the number of
non-native contacts divided by the total possible number of non-native
contacts. These two order parameters are natural for the study of
non-native interactions in protein folding. Both take on values between
zero and unity.

There are several relevant energy and entropy scales governing the
thermodynamics of folding. Let the energy of the native structure be
given by $\En$. Let the total number of contact
interactions in a fully collapsed polymer globule be given by
$M$. Asymptotically, $M$ scales like the total number of residues in
the chain, $N$, essentially because surface terms are
negligible compared to the bulk. However for a finite size system, the
mean number of
contacts per residue
(native or non-native), i.e. the coordination
number $z$, is itself a function of $N$.
We can write
the native energy as
\be
\En = M \eps = z N \eps \: ,
\label{eqEn}
\ee
where $\eps$ is then defined as
the mean native attraction energy $\eps$ ($\eps < 0$), i.e.
the native state is assumed to be fully collapsed with the
maximal number of contacts, and this is the maximal number of
total contacts of a fully collapsed polymer globule. We neglect here
the separate effects that arise from
the variance in the native interaction energies: $\delta\eps^2 = 0$.

Let the conformational entropy of an ensemble of polymer structures
characterized by the order parameters $Q$ and $A$ be given by $S_c
(Q,A)$. We can write the entropy in terms of the entropy per residue
$s_c (Q,A)$ as
\be
S_c (Q,A) = N s_c (Q,A) = M s_c (Q,A)/z \: .
\label{eqS}
\ee

In addition to the energy scales $\eps$ and $\delta\eps^2$ governing
native contacts, there
are also two energy scales governing non-native interactions. One is
the mean energy of a non-native interaction $\enn$, and the other is
the energetic variance of non-native interactions $b^2$. We keep both
of these terms, as they enter the analysis on essentially the same
footing. For configurations with $M A$ non-native contacts, the
total non-native energy is taken to be Gaussianly distributed with mean
$M A \, \enn$ and variance  $M A \, b^2$. Both of these terms contribute to
the overall ruggedness of the energy landscape by favoring non-native
configurations.

The strength of non-native interactions is taken
to be weak, so that
\alpheqn
\begin{eqnarray}
b/\eps << 1
\label{lesss1} \\
\enn/\eps << 1
\label{lesss2}
\end{eqnarray}
\reseteqn
are both satisfied. Condition (\ref{lesss1}) implies that
the ratio of the folding transition temperature $\Tf$ to thermodynamic
glass temperature $\Tg$ is large~\cite{GoldsteinRA-AMH-92}
\be
\Tf/\Tg >> 1 \: ,
\label{tftg}
\ee
i.e. the proteins we consider are strongly (but not infinitely)
unfrustrated- we are perturbing away from the G\={o} model.
Condition (\ref{lesss2}) implies that collapse and folding occur
concurrently~\cite{Thirumalai96prl}, i.e.
\be
\Tf/T_{\theta} >> 1 \: ,
\label{tfttheta}
\ee
where $T_{\theta}$ is the temperature below which non-native states
tend to be collapsed.
For a given choice of non-native interaction energies, the energies of
configurations for the ensemble of states characterized by $(Q,A)$ is
assumed Gaussianly distributed with a mean of $Q M \eps + A M \enn$ and a
variance of $A M b^2$. Then the extensive part of the log number of states
having energy $E$ and order parameters $(Q,A)$ is given by
\be
\log n(E,Q,A) = S_c(Q,A) - \frac{\left[E-\left(Q M \eps + A M \enn
    \right)\right]^2}{2 A M b^2} \: .
\label{ne}
\ee
From the definition of equilibrium temperature $T^{-1} = \D S/\D E$, one can then
find the thermal energy, entropy, and free
energy, which are given by (in units where $\kboltz=1$):
\alpheqn
\begin{eqnarray}
\frac{E(Q,A,T)}{M} &=& \eps\, Q + \left(\enn - \frac{b^2}{T}\right) A  \label{eq:E} \\
\frac{S(Q,A,T)}{M} &=& \frac{s_c (Q,A)}{z} -  \left(\frac{b^2}{2 T^2} \right) \: A \label{eq:S} \\
\frac{F(Q,A,T)}{M} &=& \eps\, Q  - T \frac{s_c(Q,A)}{z} + \left(\enn -
\frac{b^2}{2 T}\right) A \label{eq:F} \: .
\end{eqnarray}
\reseteqn
These expressions can be understood straightforwardly. In the absence
of non-native interactions ($\enn=b=0$), the thermal energy is just
the energy of native contacts times the number of native contacts, and
the entropy is just the
configurational entropy. When non-native energies are present, just as
$\eps$ couples the order parameter $Q$, so does $\enn$ couple the
order parameter $A$. When non-native energies have a variance, the
lower energy conformations (with stronger non-native contacts) tend to
be thermally occupied. This is why $\enn$ and $-b^2/T$ enter on the
same footing in the energy. The fact that the system spends more time
in fewer states means that the thermal entropy is reduced. However the
entropy (times temperature) is only reduced by half as much as the
energy, so there is a residual contribution to the free energy $E-
TS$ due to the variance of non-native interactions.

A plot of the free energy at the folding temperature of the G\={o}
model $\Tfo$ as a function of
$(Q,A)$ is shown in first row of figure~\ref{esf}, 
for equation~(\ref{eq:F})
together with the analytical model of the
configurational entropy $S_c(Q,A)$ described below.

Figure~\ref{esf} also shows plots of $E(Q,A)$, $S(Q,A)$, and $F(Q,A)$,
as well as the number of states at energy $E$,
taken from the simulation data for the off-lattice model (see section~\ref{sec:thermo}).
Plots are at the
folding temperature $\Tfo$ of the G\={o} model, for several different
values of $b$ indicated.

\subsection{Conformational entropy in terms of packing fraction}
\label{sec:entropy}

The fraction of non-native contacts $A$ is not independent of $Q$.
As more native interactions are present, less non-native interactions
are allowable, and eventually there can be no non-native contacts in the native
structure. Previous studies that investigated the folding rate at
fixed stability have explicitly included this
$Q$-dependence in equation~(\ref{eq:F})~\cite{PlotkinSS01:prot}. Here our intention is to plot
the folding rate at fixed temperature rather than at fixed
stability. For this purpose it is formally more convenient to keep
this $Q$-dependence implicit in $A$. Again this manifests itself only
as a region of allowed values of $(Q,A)$, which can be seen in
figure~\ref{esf}.

The entropy loss due to native contacts is of a different functional
form than the entropy loss due to
non-native contacts. The entropy loss due to native contacts arises
from a specific set of polymeric constraints. The entropy loss due
to non-native contact formation arises from an increase in polymer
density, a non-specific constraint. There are many collapsed unfolded
states with non-native interactions present, but only one folded state
(neglecting the much smaller entropy due to native conformational fluctuations).

We note that the conformational entropy $S_c(Q,A)$ takes into account the
extent to which polymer configurations tend to have residue pairs in
proximity, such that if they interacted, that interaction would be
considered a non-native contact. However the strength of the typical non-native
interaction ($\sim \enn \pm b$) is controlled by 2 free parameters in
the theory. When both $\enn$ and $b$ are set to
zero, the thermal entropy reduces to that in the putative G\={o}
model, with the configurational entropy $S_c(Q,A)$ remaining
unchanged.

The $A$-dependence in $S(Q,A)$ is related to the physics of collapse,
since at a given value of $Q$, the fraction of non-native contacts $A$
depends on the packing fraction $\eta$ of non-native polymer.
When $M
Q$ native contacts are
present, $\Amax \equiv M(1-Q)$ non-native contacts are allowable, and
$\Amax$ non-native contacts are present when $\eta=1$.

As detailed in Appendix~\ref{sect:entropy}, a mean field approximation
allows one to
estimate the conformational entropy $S_c(Q,\eta)$ of a disordered
polymer at $Q$ with packing fraction $\eta$ as:
\bea
S_c (Q,\eta) &=& N(1-Q) \left\{
\ln \frac{\nu}{e} - \left(
\frac{1-\eta}{\eta}\right) \ln \left( 1-\eta \right)
- \frac{1}{6} \left[
\left(\frac{\etabar (Q)}{\eta}\right)^{2/3} - 1\right]^2
\right\} \nonumber \\
&\equiv& N (1-Q) \, s_{nn}(Q,\eta) \: .
\label{eq:SQeta_bis}
\eea
Here $\etabar(Q) = \lbar(Q)^{-1/2}=[\nL(Q)/N(1-Q)]^{1/2}$, where
$\lbar$ is the mean loop length formed by native contacts at $Q$
(see equation~(\ref{eqlstar})), and $\nL(Q)$ is
the total number of loops at $Q$ (equation~(\ref{nL})).
In equation~(\ref{eq:SQeta_bis}) the quantity in curly brackets is the entropy
per residue for the remaining disordered polymer at $Q$.

Figure~\ref{seta} shows a plot of the entropy per disordered residue
at $Q$, $s_{nn}(Q,\eta)= S(Q,\eta)/N(1-Q)$, as a function of $\eta$,
for various values of $Q$. This shows that the non-native polymer
density where most of the states are (where $s_{nn}(\eta)$ is maximal) is an increasing function of
nativeness $Q$.

From equations~(\ref{eqS}) and~(\ref{eq:SQeta_bis})
\be
s_c(Q,A) = (1-Q) \, s_{nn} (Q,A/(1-Q))
\ee
The entropy per residue $s_c(Q,A)$ in equation~(\ref{eq:S}) is
then obtained from equation~(\ref{eq:SQeta_bis}) using
\be
\left. s_c(Q,A) = s_c \left(Q,\eta\right)\right|_{\eta=A/(1-Q)} \:
\label{s_per_res}
\ee
(see equation~(\ref{Scollapse}).
The free energy surface on which dynamics occurs can then be obtained from
equation~(\ref{eq:F}), and is plotted in first row of figure~\ref{esf}.  This is the
reaction surface for the coordinates $(Q,A)$.

\subsection{Effect of non-native interactions on free energy barrier
and folding rate}
\label{sec:rate_theory}

In the G\={o} model, non-native contacts are given coupling energies of
zero.
The G\={o} folding temperature $\Tfo$ is taken to be the temperature where the
unfolded and folded thermodynamic states have equal probability. This
is given through equation~(\ref{eq:F}) when $F(0,A) \approx F(1,0)$ and
$\enn = b^2=0$. We are taking $Q\approx 0$ in the unfolded state and
$A=0$ in the folded state (see figure~\ref{esf}). This yields a
G\={o} folding temperature of
\be
\Tfo = \frac{ z | \eps |}{s_c(0,\Ast(0))}
\label{tfo}
\ee
where $\Ast(Q)$ is the most probable value of $A$ at a given $Q$, as
determined below.

When considering the simulation data,  the
folding temperature is taken to be the temperature in the G\={o} model
where the unfolded and
folded thermodynamic minima have equal free energies (these minima
need not be precisely at $Q=0$ and $Q=1$).

The most
probable value of $A$ at a given $Q$ for a protein in thermal
equilibrium, $\Ast(Q)$, is obtained from
\be
\left. \frac{\D F(Q,A)}{\D A^{\ast}}\right|_Q = 0 \: .
\label{minfreeA}
\ee
Using equations~(\ref{eq:F}) and (\ref{eq:SQeta_bis}) this gives:
\be
\frac{\D s_{nn}(Q,\etast)}{\D \eta} = \frac{z \enn}{T} - \frac{z b^2}{2 T^2} \: .
\label{etastq}
\ee
where $\etast(Q)$ is the most probable packing fraction at a given value of $Q$.

Using the following definitions:
\bea
\Del \Ast(Q) &\equiv& \Ast(Q)-\Ast(0)\, ,\nonumber \\
\Del s_{nn}(Q) &\equiv& s_{nn}(Q,\Ast(Q)) - s_{nn}(0,\Ast(0))
\, , \nonumber
\eea
the minimal free energy
at $Q$, $F(Q,\Ast(Q))$, relative to the minimal free energy
$F(0,\Ast(0))$ in the unfolded state, is obtained from
equation~(\ref{eq:F}):
\bea
\Del F(Q,T) &\equiv& F(Q,\Ast(Q),T) - F(0,\Ast(0),T) \nonumber \\&&\\
\frac{\Del F(Q,T)}{M} &=& Q\left(\eps + \frac{T s_{nn}(0,\Ast(0))}{z} \right) - \frac{T (1-Q) \Del
s_{nn}(Q)}{z} + \left(\enn - \frac{b^2}{2 T} \right) \Del \Ast (Q)
\label{eqdf}
\eea

With the temperature set to the G\={o} transition temperature $\Tfo$,
the first term in brackets in equation~(\ref{eqdf}) vanishes.
The free energy barrier (over $\Tfo$) at the G\={o} transition
temperature can then be written as
\be
\frac{\Del F^{\neq}}{\Tfo}
= \frac{\Del F^{o\,\neq}}{\Tfo} + M \left( \frac{\enn}{\Tfo}
-\frac{b^2}{2\Tfosq}\right)  \Del \Ast(Q^{\neq})
\label{dftf}
\ee
where $\Del F^{o\,\neq}$ is the barrier height at $\Tfo$ with $\enn =b^2 = 0$,
i.e. the putative G\={o} barrier height, and is
given by:
\be
\frac{\Del F^{o\,\neq}}{\Tfo} = N (1-Q^{\neq}) \left(-\Del s_{nn}
(Q^{\neq}) \right)
\label{fatgo}
\ee
where the saddle point is located at $(Q^{\neq}, A^{\neq}\equiv
A^{\ast}(Q^{\neq}))$. Note that $\Del s_{nn}
(Q) < 0$ because disordered polymer dressing larger native cores is
more collapsed than that for smaller native cores. One can see
that the barriers scale extensively as a result of the mean field
approximations made above.

So we see from equation~(\ref{dftf}) that the folding barrier lowers with
increasing non-native interaction strength, namely if $\enn <0$
($b^2>0$ always),  so long as $\Del \Ast(Q^{\neq}) \equiv \Del A^{\neq} >0$.
So now
we investigate the conditions for which $\Del A^{\neq} > 0$.

From equation~(\ref{eqAeta}), the condition $\Del A^{\neq} > 0$ is equivalent to
\be
\Del A^{\neq} = \etast(Q^{\neq}) (1-Q^{\neq}) - \eta(0) > 0
\label{etagtzero}
\ee
where $\etast$ is determined from equation~(\ref{etastq}).

We are interested in the effect on the barrier when non-native
interactions are imagined to initially increase from zero.
For $\enn, b \approx 0$, the most probable packing fraction is
interpreted geometrically through equation~(\ref{etastq})
as the value of $\eta$ where the
entropy per disordered residue is maximal, i.e. the maximum of the
curves in figure~\ref{seta}. When $\enn <0 $ and/or $b>0$, $\etast$ is
determined as the value of $\eta$ slightly to the right of the
maximum in the curves in figure~(\ref{seta}). The most probable packing fraction
as a function of $Q$ is plotted in figure~\ref{figetaq}.

Equation~(\ref{etagtzero})  is not
a particularly robust condition.
While $\etast(Q)$ is certainly a monotonically increasing function of $Q$
as can be seen from figure~\ref{figetaq}, the factor of $(1-Q)$ in
equation~(\ref{etagtzero}) de-emphasizes, or may reverse, the trend in
$\Ast(Q)$. In the earlier work addressing the trend in rates at {\it
  fixed stability} rather than {\it fixed temperature}, the
factor determining whether rates would increase was merely the increase in
packing fraction $\Del \eta(Q)$ by itself~\cite{PlotkinSS01:prot}.

The derivative of $s_{nn}$ in equation~(\ref{etastq}) can be
straightforwardly determined from equation~(\ref{eq:SQeta_bis}), and
equation~(\ref{etastq}) then becomes a non-linear equation for $\etast(Q)$
that can be solved numerically. The result is shown in
figure~\ref{figetaq}. The packing fraction increases as the length of
disordered loops becomes shorter (inset of
fig~\ref{figetaq}), and thus increases monotonically with
nativeness $Q$.

Once $\etast(Q)$ is known, $\Del \Ast(Q)$ can be obtained
from~(\ref{etagtzero}).
This determines the trend in the barrier height by
equation~(\ref{dftf}). A plot of $\Del \Ast(Q)$ is shown in
figure~\ref{delaq}. We can see that if the barrier position $Q^{\neq}$
resides in a window of $Q$ where $\Del A(Q) >0$, the barrier decreases
with increasing non-native interaction strength, {\it for weak
  non-native interactions}. Otherwise the barrier
increases with increasing non-native interaction strength.

When non-native interactions are weak, the folding kinetics are single
exponential:
\be
\kF = k_o (\enn, b) \e^{-\Del F^{\neq} (\eps,\enn,b)/T} \: .
\label{eq:rate}
\ee
Increasing the strength of non-native interactions slows the prefactor $k_o$,
due to the effects of transient trapping. However as $\enn$ and $b$
are increased from zero, this slowing effect on $k_o$ does not become
significant until a non-zero characteristic value, which would indicate
the onset of a dynamic glass transition in an infinite sized
system (see~\cite{WangPlot97,TakadaS97:pnas,PlotkinSS01:prot,EastwoodMP01} for more
detailed treatments of this effect). In a finite
system the activation time
$\sim k_o^{-1}$ increases dramatically but only when $b> \bA$ or
$\enn > \enn^{\mbox{\tiny A}}$. The values of the energy scales $\bA$ and
$\enn^{\mbox{\tiny A}}$
are of order $T$, so there is a fairly large window upon increasing
$b,\enn$ from zero where the prefactor $k_o$ is unaffected to the
first approximation. In this regime the effects on rate are governed
solely by the effects on barrier height. Hence the decrease in
barrier height shown above as $\enn,b$ are increased from zero may be
associated with an increase in folding rate.

In the next section we test the theoretical prediction directly with
simulations of a model protein. The upshot is shown in
figures~\ref{fig_rate}~(b) and~(c) below, which show indeed an
increase in folding rate with increasing non-native interaction
strength. 

\section{Comparison between theoretical prediction and simulation results}
\label{sec:simulations}

We have thoroughly explored the range of validity of the approximations made
in the analytic theory by comparing the predictions with the results
obtained from simulations on a G\={o} model increasingly perturbed by
the addition of non-native interactions (see
Appendix~\ref{sect:app-sim} for details on simulation).

A close and quantitative comparison of the results from theory and simulations
is possible if corresponding thermodynamic quantities and reaction coordinates
are identified.
For this purpose, before we proceed to test the prediction on rate enhancement,
three main points of the theory have to be examined in comparison with the results
from simulations:

$\bullet$
definition of the reaction coordinates $Q$ and $A$

$\bullet$
allowed values of the reaction coordinates (\ie correlation between $Q$ and $A$)

$\bullet$
approximations made in the definition of energy and entropy as functions of the
reaction coordinates

These points are clearly interconnected and all effect the detailed shape of the
free energy landscape, the value of the folding temperature, and the identification
of the folded, unfolded, and transition state ensembles.
We expect that the assumptions we have made in the analytical theory
do not qualitatively change the theoretical predictions, nevertheless a careful
dissection of the basic ingredients we have used is needed for a  quantitative
assessment of the results.

In the following we discuss in detail each of the points above.
Unless otherwise specified the following results are all obtained from simulations at the G\=o
folding temperature $T_f^0$, for all values of $b/\epsilon$.

\subsection{Definition of reaction coordinates}
\label{sec:Asim}

The reaction coordinate $Q$, defined as the fraction of
native contacts formed in a given protein configuration, is readily associated
to configurations sampled by simulations (see Appendix~\ref{sect:app-sim}).
More care has to be used in transposing the other reaction coordinate we
have used in the theory, $A$ (defined as the fraction of non-native
contacts formed), to the analysis of simulations data.
In the analytical theory we have assumed that the maximum number of non-native
contacts that can be formed at a certain stage of the folding reaction
does not depend on the perturbation strength, and is a function of the degree of
nativeness, $Q$, that is $A_{max}(Q) = 1- Q,\;  \forall b$ (see equation (\ref{eqAeta})).
This implies that no non-native contacts can be formed in the native state
($A_{max}\sim 0$ if $Q\sim 1$), and {\it vice-versa} ($A_{max}\sim 1$ if $Q\sim 0$).
This assumption in the theory allows us to simplify the analytical calculations
but does not qualitatively affect the results. The dependence on $Q$
of the maximum number
of non-native contacts can be directly checked in simulations.
In this regard, an important difference between theory and simulations is
that a certain number (typically $\sim$ 5) of non-native contacts
can be accommodated in a protein configuration with $Q\sim 1$ and minimal
(less then 1\;\AA\,) rms deviation from the pdb native structure.
The increased number of contacts
around the native configurations
arises mainly from the fact that native or non-native contacts are
considered formed in a small but finite
length range (typically $\sim$ 1\AA\,) around the minimum of the interaction
potential. This leads to probable formation of some non-native
contacts as the protein undergoes fluctuations around the native state.

Figure~\ref{fig_nnformed} shows that a subset of 6 non-native contacts
is formed with probability > 0.25 in the native state ensemble for
$b/\epsilon=1.3$. Similar results are obtained for
all values of $b/\epsilon$ used in this study, although the particular
set of non-native contacts formed in each case depends on the choice of non-native
interactions (data not shown).

Contacts that are easily formed in the native state can not be
considered non-native, even when they are not listed as native contacts
in the unperturbed G\=o-like Hamiltonian. In fact, contacts that can be
made in the native state are not competing against the formation of the
native structure, rather they are assisting it. In order to remove this effect, we
introduce a new reaction coordinate $A^{\prime}$, defined as the fraction
of non-native contacts formed, restricting the list of non-native interactions
only to the ones with a probability of contact formation in the native state
ensemble smaller than a cutoff value $p_{c}$. The native ensemble for each
sequence is identified as all configurations with $Q > 0.9$ sampled in simulation
for that sequence.
The results presented in the following are all obtained with a probability cutoff
$p_{c}=0.1$. Smaller values of $p_{c}$ yield essentially the same results.
The reaction coordinate $A^{\prime}$ is then used in this study to compare results
from simulation with the theoretical predictions.

Another approximation that can be directly checked in simulation is on the
maximum number of non-native contacts that can be formed at different stages of
the folding reaction. In the analytical theory, the fraction of non-native contacts, $A$,
is a function of the fraction of native contacts formed in a configuration, $Q$, and
of the packing fraction $\eta$ of the non-native part of the protein: $MA = \eta(1-Q)M$
(see equation~(\ref{eqAeta})), with $\;0 \leq \eta \leq \ 1,\; \forall Q\;$.
The maximum number of non-native contacts is then $A_{max} M = (1-Q)M$, and
the maximum total fraction of all contacts (native and non-native) is $(A+Q)_{max} = 1, \forall Q$.
Indeed, the maximum number of all contacts (both native and non-native) recorded in simulations is
close to the number of native contacts formed in the native state, \ie $M(Q+A^{\prime})_{max}\simeq M$,
for all values of the parameter $b$ examined in this study (see figure~\ref{fig_Amax}(a)).
Figure \ref{fig_Amax}(b) shows the behavior of the average number of non-native contacts  formed in
simulation (both coordinates $A$ and $A^{\prime}$ are plotted), as a function of $Q$, for a
perturbation $b/\epsilon = 0.5$ (right panel), and the value of $Q$ corresponding to the maximum
of $\langle A^{\prime}\rangle$ (the corresponding $Q$ for the uncorrected coordinate
$\langle A\rangle$ is also shown).
Interestingly, the peak in the average number of non-native contacts is detected
for a value of $Q$ corresponding to a pre-transition state stage of the folding. A pre-TS peak
is observed in both theory and simulations, although in the theory it is closer to the unfolded
state than what detected in simulations (see figures \ref{fig_Amax}(b) and \ref{fig_Asim}(a)).

Figures \ref{fig_Amax}(a)-(b) and \ref{fig_Asim}(a)-(b) present a thorough comparison between
the allowed and most probable values for the fraction of non-native contacts at different
stages of the folding, as obtained from  theory and simulations. Although the maximum number of
non-native contacts is always detected in a pre-TS region, independently on the value of $b/\epsilon$,
it is clear from \ref{fig_Amax}(a) and \ref{fig_Asim}(a) that larger values of $b/\epsilon$ yield
larger a number of non-native contacts formed, particularly in the
unfolded ensemble. Interestingly however,
the number of non-native interactions rapidly decreases to zero in
region with very small $Q$. The cause of this effect is not contained
in the analytical expressions~(\ref{eq:F}), where it is assumed that
$A_{max} = 1-Q$. This result is due partly to
coupling between non-native contacts and the
angle and dihedral terms in the simulation Hamiltonian (which are not present in
the theory). This is a finite size effect which tends to increase the
polymer stiffness relative to that in the theory,
which used a bulk approximation for thermodynamic quantities. Compact
states with $\eta \sim 1$ in which only non-native interactions are
present have large energetic cost and are formed very rarely. Another
source of this effect is that forming collapsed conformations induces some native
contacts to be formed, due to the finite range of interactions. This
effect is particularly important for short-range contacts among
residues closely separated in sequence, and does not necessarily go
away as one considers larger size systems.
This is the complementary effect to the already
mentioned fact that in simulations non-native interactions are formed in the native state
(that has led us to a redefinition of the simulation reaction coordinate $A^{\prime}$).

In order to quantify this effect we have generated a large (50) set of non-native energy
distributions with high and very high variance ($b/\epsilon \geq 2$ and $b/\epsilon \gg 2$).
Sequences with these high values of $b/\epsilon$ are not able to fold to the selected native
structure, but are useful to explore the region of the configurational space corresponding to
compact structures with the maximum number of non-native contacts formed.
We expect the glass temperature of these sequences to be higher than their folding temperature
(see next section).
After an initialization at very high temperature ($T \gg T_f^0$), a
large number ( $> 1000$) of quenching simulations ($T \ll T_f^0$) has been performed for each
sequence to generate a representative sample of compact misfolded structures. The maximum
fraction of non-native contacts that can be formed $A^{\prime}_{max}$ is thus defined as the
largest values of $A^{\prime}$ among the vast pool of structures obtained by adding the results
from the quenching simulations for high $b/\epsilon$ values to all configurations collected in
simulations at any temperature and for any value of $b/\epsilon$ used in this study.
Figure \ref{fig_Asim}(b) shows the behavior of $A^{\prime}_{max}$ as a function of the fraction
of native contacts present in the structures. The theoretical assumption on the maximum fraction of
non-native contacts $A_{max} = 1-Q$ holds remarkably well up to the values of $Q \lesssim 0.15$, that
corresponds to the unfolded state minimum in the free energy landscape (see figure~\ref{esf}).
From these results we then expect the unfolded region of a free energy landscape associated with
the simulated protein Hamiltonian to be somewhat compressed
toward smaller values of $A$ with respect to the theoretical prediction.

\subsection{Energy, entropy and free energy landscape}
\label{sec:thermo}

Figure~\ref{esf} presents the energy, entropy, and  free energy profiles obtained from simulations,
as a function of
the reaction coordinates $Q$ and $A^{\prime}$, for three different values of the perturbation
parameter $b/\epsilon$. The corresponding quantities obtained from the analytical theory, with
all the parameters set equal to the simulations parameters
(i.e. rightmost column in table I) and
$b = 0.3\epsilon$, is also shown for comparison.
For a more direct comparison with the results from simulations, the thermodynamic
quantities from theory are only plotted in regions populated with probability larger than
$2\times 10^{-7}$, as we have typical samplings of $\sim 5\times 10^{6}$ configurations in
folding/unfolding simulations.
It is apparent from figure~\ref{esf} that the region of the $(Q,A^{\prime})$ space populated
with high probability in simulations differs somewhat from the $(Q,A)$ region predicted by theory.
Several factors are responsible for this difference and have to be considered before one
tests the predictive power of the theory with the simulation results:\\
(i) The unperturbed energy function used in simulation includes a self-avoiding term for all
non-native contacts, that is maintained in the perturbed Hamiltonian
(see equations~(\ref{go_ham}),~(\ref{expr_nn}) and figure~\ref{fig_lj}).
This energy term is not explicitly considered in the theory. The short-distance repulsive
interactions limit the formation of non-native contacts (especially for small values of
$b/\epsilon$), and shifts the most populated regions of the folding landscape toward lower
values of $A$. This effect also accounts for most of the differences in the energy landscape
between theory and simulation results (see figure ~\ref{esf}). \\
The analytic expressions are obtained in the thermodynamic limit, while simulations are
performed for a small protein (57 residues). The theoretical expressions do not
explicitly keep track of finite-size effects due to polymer stiffness. However the extra
effects of polymer stiffness seen in the simulations only enhances the
theoretically predicted rate acceleration effect (see section~\ref{sec:rate}). \\
(iii) The functional form for the entropy is approximated in the theory and it is not expected to
quantitatively reproduce the simulation results exactly. Particularly, the theoretical assumption on the
allowed values of $A$ at different $Q$ (\ie equation~\ref{eqAeta}) directly enters the derivation
of the entropy (see Appendix~(\ref{sect:entropy}) for details), and contributes to the
relative "distortion" of the theoretical free energy landscape with
respect to the landscape in the simulations. Nevertheless,
the overall qualitative behavior of the entropy is correctly captured by the theory (see third
column of figure ~\ref{esf}).\\
(iv) The position of the folded and unfolded free energy minima emerging from simulation data
differs from $Q=0$ and $Q=1$, as assumed in the theory (see also section \S~\ref{sec:rate_theory}).

Overall, the destabilization of the folding free energy landscape upon introduction of non-native
energy perturbation is strongly reduced in simulations with respect to what predicted by the theory.
In fact, while in the theory a perturbation of $b/\epsilon \sim 1$
renders a protein unfoldable (\ie $T_f/T_g\sim 1$, see
equation~(\ref{t_glass}) and ref.~\cite{PlotkinSS01:prot}),
it is found in simulations that all sequences generated with a perturbation parameter
$b/\epsilon \lesssim 1.7$ (entering the Hamiltonian~(\ref{expr_nn})) are able to reversibly fold/unfold
at the G\=o transition temperature $T_f^0$. The next section quantifies this difference in the
destabilization of the folding mechanism by comparing the folding and glass temperatures
computed in simulations with their corresponding theoretical predictions.

\subsection{Folding temperature and glass temperature}
\label{sec:temp}

The folding temperature $T_f$ of each protein model is estimated in simulation as
the temperature corresponding to the peak in the specific heat curve (see
figure~\ref{fig:cv}a). This value is in good agreement (within the error bars)
with the value obtained from the alternative definition of $T_f$ as the temperature
 at which the folding and unfolding states have the same free energy
(see figure~\ref{F_vs_Q}b).

From equation~(\ref{eq:F}), upon increasing non-native interaction
strength (increasing $b$, and/or increasing $|\enn|$ with $\enn<0$),
the free energy $F(Q)$ lowers with respect
to the G\={o} free energy at fixed temperature $T=\Tfo$. During this
change of Hamiltonian, the free
energy of the native structure remains roughly constant at $\En$ (see
figure~\ref{F_vs_Q}~(c)). Even though the unfolded state is stabilized
with respect to the folded state during the process of increasing
non-native interaction strength at fixed temperature, the folding rate nevertheless
accellerates, because the free energy of the transition state lowers
more than the unfolded state does. This is described in more detail
below, with the result shown in figure~\ref{fig_rate}~(b). 

The thermodynamic glass temperature $T_g$ can be estimated by using the
results obtained in the framework of the random energy model (REM)
\cite{Derrida81,BryngelsonJD87}.
As the energetic frustration of the system
arises from randomly assigned non-native interactions,
we assume that the energy of compact (misfolded) structures in the unfolded
ensemble is Gaussianly distributed, with mean value $\langle
E_{nn}(Q_{u})\rangle= M A_{max} \enn$  and
variance $\delta E^2_{nn}(Q_{u})=b^{2}A_{max}M$, where $MA_{max}$ is the maximum number of
non-native contacts the protein can form. In the theory the maximum number of non-native
contacts was approximated at $Q \sim 0$ as the total number of native contacts
$MA_{max} = M$. As we have already discussed in section \S~\ref{sec:Asim}, the actual
maximum number of non-native contacts detected in simulation is smaller
than the theoretical value, and it expected to (slightly) vary with different realizations of
the non-native noise (see figure~\ref{fig_Amax}).

The REM glass temperature is defined by the vanishing of the thermal
entropy~\cite{Derrida81,BryngelsonJD87}), which corresponds to setting
equation~(\ref{eq:S}) to zero:
\be
S(Q_{u},A_{max}, T_g) =  N s_c (Q_{u},A_{max}) - \frac{M A_{max}b^2}{2
  T_g^2} = 0 \: ,
\ee
however here we let $A_{max}$ be a new parameter. This gives for the
glass temperature:
\be
T_g =  b \sqrt{ \frac{z A_{max}}{2 s_c}} = \frac{\delta E_{nn}(Q_{u})}{\sqrt{2 N s_c}}
\label{t_glass}
\ee
where $\delta E_{nn}(Q_{u})
= MA_{max}b^2$ is the energetic variance over the set of misfolded structures.
%
%
For each protein model (\ie, each value of $b/\epsilon$) we have performed several
(more than 500) short quenching simulations to explore the compact configurations in the unfolded
ensemble.
A different open configuration is initially created by
means of ancillary high temperature simulations (with $ T \gg T_f$), then
rapidly quenched to very low temperatures ($ T \sim T_f/10$, $T \sim
T_f/25$, and $T \sim T_f/50$). The fluctuations of the non-native energy
in the compact misfolded  configurations recorded during the quenching simulations are used to
compute $\delta E_{nn}(Q_{u})$ entering expression~(\ref{t_glass}).

Figure~\ref{fig_glass}(a) shows the folding temperature $T_f$ and the
glass temperature $T_g$ obtained from simulation, as a function of the strength of the
non-native energy perturbation, $b$ (in units of the native energy per contact, $\epsilon$). The
folding temperature is almost constant in the range shown, while the glass temperature raises
from zero ($b = 0$ corresponds to the plain G\=o-like model with no energetic frustration,
see equation~(\ref{go_ham})), to values close to $T_f$ for large non-native perturbations
($b \gtrsim 1.6$). When $T_g/T_f \approx 1$ many low energy misfolded structures compete with the
native state and folding is dramatically slowed down. As the ratio $T_g/T_f$ increases
beyond unity, the system is no longer self-averaging, and different realizations
of the non-native perturbation can lead to different folding mechanisms
consistent with the same native topology. This point is discussed in a
separate publication~\cite{Plot-Clem03}.
%
%
The glass temperature predicted by the theory for different values of $b$ are also obtained
from equation~(\ref{t_glass}), with $\delta E^2_{u}= MA_{max}(Q_{u})\;$,
$\;MA_{max}(Q) = M(1-Q)$, and $Q_{u}$ corresponding to the
unfolded free energy minimum at $T_g$. The theoretical folding temperature is evaluated as
described in section~\S~\ref{sec:rate_theory}.
The comparison of the folding and glass temperatures from simulation with the corresponding
values predicted by the theory (dotted curves in figure~\ref{fig_tf}(a)) clearly shows that the
destabilizing effect of the non-native energy perturbation on the folding process (quantified
by the ratio $T_g/T_f$ is much reduced in simulation with respect to the theoretical prediction.
Each value of $b$ used in simulation ($b_{sim}$) is plotted in figure~\ref{fig_tf}(b) as
a function of the value of $b$ used in the theory ($b_{theory}$) which yields the same $T_g$.
The corresponding $T_f(b_{sim})$ \;(from simulation) and $T_f(b_{theory})$\; (from theory) are
also found equal within the error bar.

\subsection{Folding rate enhancement/depression upon non-native energy perturbation}
\label{sec:rate}

The theoretical prediction on folding rate enhancement upon small non-native energy
perturbation is expected to hold for values of $b$ with a corresponding small ratio
$T_g/T_f$. A perturbation that largely increases the ratio $T_g/T_f$
will also largely decrease the prefactor $k_0$ in
equation~(\ref{eq:rate}), and folding then slows
(see discussion in section~\S~\ref{sec:rate_theory}).
Because of the extended range of $b$ for which the condition $T_g/T_f \ll 1$ remains
valid in simulation (see previous section), we expect to detect a rate enhancement in simulation
up to values of $b\sim 1$, i.e. the theory is conservative in that
rate is enhanced over a wider range of $b$ in the simulations.

We have shown in the previous section that the analytical theory reproduces
correctly, at a qualitative level, the thermodynamics quantities measured in simulation, although we
have highlighted some quantitative differences. The effect of these differences
on equation~(\ref{dftf}) which predicts the rate enhancement is expected to be confined to
the precise evaluation of the difference in the number of non-native
contacts between the transition state and unfolded state, $\Del A^{\neq}
\equiv \Del \Ast(Q^{\neq})$, and to a lesser extent the precise positions of the transition
state and unfolded state,  $Q^{\ddag}$ and $Q_{u}$
respectively. Equation~(\ref{dftf}) can thus be directly
and quantitatively tested if $\Del \Ast(Q^{\neq})$ is evaluated from simulation.
Figure~\ref{fig_rate}(a) shows the difference $M \left(\langle A^{\prime}
\rangle_{TS} - \langle A^{\prime} \rangle_{U} \right) \equiv M \Del A'^{\neq}$ between the
average number of non-native contacts $M \langle A^{\prime} \rangle_{TS}$, formed in
the simulated transition state ensemble, and the average $M \langle A^{\prime} \rangle_{U}$,
formed in the simulated unfolded ensemble. This number slightly varies over the range of $b$
values where we expect to find the rate enhancement effect ($T_g/T_f \lesssim 0.25$ up
to $b \lesssim 1.3\epsilon$).
Since the variation of $M \Del A'^{\neq}$ with $b$ in this range is smaller than  error bar associated
to it, we consider its average over the different $b$ values (straight red line in figure ~\ref{fig_rate}(a)). This average value is then
used in equation~(\ref{dftf}); the resulting quantity $\ln (k/k_o) = (\Del
F_o^{\neq} - \Del F^{\neq})/\Tfo$ is compared
with the difference in log folding rate estimated directly from a large set of folding simulations.
Figure~\ref{fig_rate}(b) shows that the agreement between the values predicted from
equation~(\ref{dftf}) (dashed black line) and simulation results for the rate (red dots)
and barrier height (blue dots) is indeed remarkably good up to $b \lesssim 1-1.1$.

Folding rates obtained from simulations performed with $b =0$ and
variable $\enn$ are also plotted
in the figure~\ref{fig_rate}(c). 
As predicted by the theory, rates accellerate when $\enn <0$
(attractive non-native interactions) and decellerate when $\enn>0$
(repulsive non-native interactions). The theory gives excellent
agreement with the simulations in the perturbative limit (dashed line
in figure~\ref{fig_rate}(c)).
The effect on the rate (at $T_f^0$) of
a perturbation with $(b,\:\enn) =(0,\: - b^2/2 \Tfo)$ is equivalent to
the case with $(b,\:\enn) =(b,\: 0)$.
When $\enn$ becomes sufficiently attractive, the prefactor
becomes increasingly important in determining the folding rate, and
rates begin to decrease dramatically. 

\section{Conclusions}

In this paper we derived a theory for the change in the free energy
barrier height to protein folding, as the strength of non-native
interactions is varied. We find that the barrier height initially decreases as
the strength of non-native interactions increases.

This means that if one considers two idealized protein sequences,
one completely unfrustrated (a so-called G\={o}-like protein), and
one with weak non-native interactions that are either attractive or
randomly distributed,  the mildly frustrated protein will tend to fold
faster at the same temperature, particularly when the temperature
is near the transition temperature of the G\={o} protein. This result
follows from energy landscape theory~\cite{PlotkinSS01:prot}.

The criterion for the rate to increase is related to an increase in
packing fraction in the transition state relative to the unfolded
state (equation~(\ref{etagtzero})).

The rate increase is supported by the theoretical proposal that
proteins exhibit a dynamic glass transition at non-zero
temperature. The consequence of this is that the pre-factor to the
rate is initially unaffected as non-native interactions are increased
in strength from zero. Thus rate-determining effects for nearly
unfrustrated proteins arise largely
from effects on the folding barrier.

Off-lattice simulations of a coarse-grained $C_{\alpha}$ model of src-SH3
were used to test the theoretical predictions. Simulation results showed even more
robust rate-enhancement effects
than the theory, due essentially to chain stiffness and contact range
effects that decrease the number of non-native interactions in the
unfolded state.
When these corrections are included, the theory and simulations are in
very good agreement (figure~\ref{fig_rate}).

The experimental relevance of this effect (reduced number of
non-native contacts in the unfolded state) depends on whether the fraction
of native contacts formed, $Q$, is a good reaction coordinate for these systems.
For unfrustrated or nearly unfrustrated systems, $Q$ has been shown to work well
as a reaction coordinate in lattice models~\cite{NymeyerH00:pnas} (lattice
models have limited move-sets that may further hinder the use of
$Q$ as a reaction coordinate, relative to the off-lattice system
we studied here), and off-lattice G\=o-like models of short
proteins~\cite{Clementi2001:JMB,SheaJE00}.

Random non-native interactions as well as attractive non-native
interactions both speed the folding rate, when they are perturbatively
small compared to the large native interaction energies that drive
folding.  The analysis here was done at the transition temperture of
the G\={o} model. Since the coupling of collapse with folding is
fairly generic, it is expected that the effect of rate-enhancement
would also be seen at different temperatures and stabilities.

The effect of rate enhancement by non-native stabilization has been
seen in several simulation models~\cite{LiL00,FanK02,CieplakM02,TreptowWL02}, as well as
experiments
involving the strengthening of non-specific hydrophobic interactions
in $\alpha$-spectrin SH3~\cite{VigueraAR02}.

Some proteins are thought to be sufficiently frustrated that
non-native interactions may limit the folding rate. These proteins
would have non-native energy scales somewhat larger than unity in
figure~\ref{fig_rate}b, at least for some non-native contacts.
In some proteins such as Lysozyme, these
non-native interactions are thought to stabilize early-formed structures to
prevent degradation or
aggregation~\cite{Klein-SeetharamanJ02}. All-atom simulations of the
$36$-residue Villin headpiece segment suggested that the breaking of
non-native interactions incorrectly packed in the hydrophobic core may
form the rate-limiting step on some folding
trajectories~\cite{ZagrovicB02} (the authors caution however that this may
indicate frustration in Villin, or may indicate an artifact of the
force-field employed). For proteins that must escape kinetic traps to
fold, it is possible that other evolutionary mechanisms in addition to
funneling may assist folding, such as the selection for amino acids
that reduce the escape barrier from the trap~\cite{PlotkinSS03}.

To quantify the rate enhancement it was necessary to treat the entropy
of a finite-sized, self-avoiding chain -- a problem of some interest to
polymer physics. The mean-field Flory entropy of a long, self-avoiding
chain of packing fraction $\eta$ must be modified when the chain is
sufficiently short that configurations with the characteristic radius
of gyration have non-zero packing fraction. Then most states have a
finite packing fraction dependent on the length of the chain, rather
than the bulk value of zero.

From the analysis of simulation data and its comparison with the theory,
it emerges that non-native perturbations up to values of $b \sim \epsilon$
yield values of $T_g/T_f \lesssim 0.4$ (see figure~\ref{fig_tf}), that can still be considered
realistic for proteins. All sequences characterized by this range of
frustration are fast-folders, however the range of ruggedness is sufficiently wide
that a variety of scenarios are possible {\it a priori} for the
folding rate. Both rate enhancement and reduction are compatible for
realistic levels of frustration. This fact may have been exploited by natural
evolution to select different effects for different purposes (in the same
structural family). It is worth noticing that the observed rate enhancement/reduction
induced by non-native interactions is limited to less than an order of magnitude
(at least for the SH3 fold considered here), thus it can not be used to explain the
much larger variation (spanning more than 6 orders of magnitude) of folding rates
experimentally observed for single-domain, two-state folding proteins~\cite{Plaxco98,PlaxcoKW00:biochem}.

In this paper we made a very simple generalization of the G\={o}
Hamiltonian for a foldable protein, and found this resulted in
non-trivial and rich behavior of the dynamics of the system.
It will be interesting to see what new phenomena emerge from further
considerations of the Hamiltonian describing biomolecular folding and
function.

\section{acknowledgments}

We express our gratitude to Jos\'{e} Onuchic for numerous insightful
discussions and support.  The preliminary stage of this work has been
funded by NSF Grants $9603839$, $0084797$, NSF Bio-Informatics
fellowship DBI9974199, and the La Jolla Interfaces in Science program
(sponsored by the Burroughs Wellcome Fund).  C.C. acknowledges funds
from the Welch foundation (Norman-Hackerman young investigator award),
start-up funds provided by Rice University, and Giovanni Fossati for
suggestions and continuous encouragement.  S.S.P. acknowledges funding
from the Natural Sciences and Engineering Research Council, start-up
funds from the University of British Columbia, and the Canada Research
Chairs program. Members of Clementi's group and Plotkin's group
are warmly acknowledged for stimulating discussions.

\renewcommand{\theequation}{\thesection.\arabic{equation}}
\renewcommand{\thesection}{\Alph{section}}
\setcounter{section}{0}

\appendix

\section{Entropy of a partially collapsed protein as a function of the number of native
and non-native contacts}
\label{sect:entropy}

In terms of the packing fraction the total number of non-native contacts is
\be
M A = M \eta (1-Q) \: ,
\label{eqAeta}
\ee
where $\eta$ is the packing fraction of non-native polymer surrounding
the dense ($\eta=1$) native core.

The mean-field configurational entropy of a self-avoiding polymer of
$n$ links with packing
fraction $\eta$ is given by~\cite{FloryPJ53,SanchezIC79}
\be
\frac{S_c^{\mbox{\tiny{SA}}}(\eta)}{n} = \ln \frac{\nu}{\e} - \left(
\frac{1-\eta}{\eta}\right) \ln \left( 1-\eta \right)
\label{Scollapse1}
\ee
The conformational entropy of the self-avoiding walk in terms of the
fraction of non-native contacts $A$ is given by
\be
S_c^{\mbox{\tiny{SA}}}(A) =
\left. S_c^{\mbox{\tiny{SA}}}(\eta)\right|_{\eta=A/(1-Q)} \: .
\label{Scollapse}
\ee
Expressions~(\ref{Scollapse1}) and~(\ref{Scollapse}) imply that the polymer
chain in question will tend to have $\eta=0$ and $A=0$ since
this maximizes the entropy. However a finite-length chain of $n$ links
tends to have a non-zero packing fraction given by
\be
\eta(n) \approx \frac{n a^3}{R_g(n)^3} \approx \frac{n a^3}{\Delta R^3}
\label{etadef}
\ee
where $a^3$ is the
volume per monomer and $R_g$ is the radius of gyration of the
chain. Up to factors of order unity the RMS size of the polymer can be
used as well.
For chains obeying ideal statistics $\eta(n) \approx
n^{-1/2}$. For self-avoiding chains in a good solvent, accounting for
swelling gives $\eta(n) \approx n^{-4/5}$. However these expressions for
the typical packing fraction are {\it inconsistent} with
expression~(\ref{Scollapse1}), which implicitly assumes an infinite
chain limit. For finite-length chains, we seek an entropy
function which is peaked at non-zero values of $\eta$.

The assumption of ideal chain statistics for protein
segments is not as bad as it may at first seem, because disordered polymer
segments interact with each other in addition to themselves.
Polymers in a melt obey Gaussian statistics~\cite{amorphous:macsci}.
Swelling due to excluded volume is counterbalanced by compression due
to the surrounding polymer medium if the protein is sufficiently
large. However, for polymer loops dressing a native core,
self-avoidance must be taken into account to fully treat the effects
of non-native interactions.

We take the effects of self-avoidance, finite size, and ``inter-loop''
interactions into account by letting the
number of walks with density $\eta$ be the number of states at density
$\eta$, $\exp S_c(\eta)$ above, times the probability that an ideal walk of
$\ell$ steps has density $\eta$:
\be
\Omega(\eta, \ell) = \e^{S_c(\eta,\ell)}  \, p(\eta|\ell) \: .
\label{eqneta}
\ee

For smaller values of $\ell$, larger values
of $\eta$ are more probable. But at higher values of $Q$, smaller values
of $\ell$ are more probable. Hence the non-native packing fraction tends to
increase with folding. This is the effect we are quantifying here.

The number of states of the disordered polymer with packing fraction
$\eta$, at degree of nativeness $Q$, is given by
\be
\Omega(\eta,Q) = \prod_{\ell}  \: \Omega(\eta, \ell) \, n(\ell|Q) =
\prod_{\ell}  \: \e^{S_c(\eta,\ell)}\,
p(\eta|\ell) \, n(\ell|Q) \: .
\label{totnumberstates}
\ee
This is the product over all lengths $\ell$, of the
number of states for a loop of length
$\ell$ and packing fraction $\eta$, times the probability that the
loop of finite length $\ell$ has packing fraction $\eta$, times the number of
disordered loops of length $\ell$ at nativeness $Q$.

We now seek the probability distribution $p(\eta|N)$. Consider for the
moment one dimensional random walks of $N$ steps, which we generalize
to three dimensions below.
The probability $p(\eta|N)$ is maximal at the value of $\eta$
corresponding to a Gaussian distribution for the chain
(i.e. $N^{-1/2}$ above). Again however, this alone does not account for
self-avoidance, which is why $S_c(\eta,\ell)$ must be included later
in the analysis. If we let the fraction of walks with variance
$\lambda N a^2$ by given by $p(\lambda| N)$, the problem of finding
$p(\eta|N)$ is equivalent to the problem of finding $p(\l|N)$. This is
the probability a walk of $N$ steps has an anomalous
variance of $\l N
a^2$, given that the most-probable distribution of walks $\pbar$ is given by
\be
\pbar (x)  = (2 \pi N a^2)^{-1/2} \, \exp\left(-\frac{x^2}{2 N
  a^2}\right) \: .
\ee

The probability $p(\l,N)$ can be written as a functional integral over
all possible probability distributions, of the probability of a given
distribution $P[ p(x) ]$, times a delta function which counts only
those walks that have a given variance of $\l N a^2$:
\be
p(\l| N) = \int {\cal D} p(x) \: P[p(x) ] \:  \d \left( \l  -
\frac{1}{N a^2}\int \! dx \: x^2 p(x)\right) \: .
\label{functn2}
\ee
The calculation is performed in \S~\ref{sect:pln}. The result
for the probability distribution of anomalous variance $\l$ is:
\be
p(\l | N) = \sqrt{\frac{N}{6\pi}}\: \e^{- N\left(\l -1\right)^2/6} \:
\label{pln}
\ee

We can see from equation~(\ref{pln}) that the mean value of $\l=1$, meaning
that a walk of $N$ steps has on average a variance $N a^2$. However
there is variance $\delta \l^2 = 6/N$ in the distribution, so that
some walks are either particularly diffuse or condensed
statistically. The anomalous variance
decreases monotonically with increasing $N$.

For a walk in three-dimensions, we define $\l$ through the variance
\be
\Delta {\bf R}^2 = \l N a^2 \: .
\label{delR-lamb}
\ee
From the definition of $\eta$ in
equation~(\ref{etadef}), the parameter $\l$
depends on $\eta$ (and $N$) as
\be
\l(\eta) = \eta^{-2/3} N^{-1/3}
\ee

The probability
distribution of walks of density $\eta$ is then given by
\be
p(\eta|N) = p(\l(\eta)|N) \left| \frac{d\l}{d\eta} \right|
\label{pwjacob}
\ee
(the Jacobian is not particularly important here as it enters the
entropy only logarithmically).

With the above definition in equation~(\ref{delR-lamb}) for $\l$ in
three-dimensions, $p(\l |N)$ remains unchanged from the
one-dimensional form in equation~(\ref{pln})
(see Appendix~\ref{sect:pln}).

The conformational entropy for a chain of length $\ell$ having packing
fraction $\eta$ is
obtained from equations~(\ref{Scollapse1}),
(\ref{eqneta}),(\ref{pln}), and
(\ref{pwjacob}):
\be
S(\eta,\ell) = \ln \Omega(\eta,\ell) \approx
S_c(\eta,\ell)  - \frac{\ell}{6} \left[
\left(\frac{\etabar}{\eta}\right)^{2/3} - 1\right]^2
\label{eqsceta}
\ee
where $\etabar = \ell^{-1/2}$ gives the most probable value for the
packing fraction for an ideal (non-self-avoiding) chain of length $\ell$.
For an interacting chain, enthalpy and entropy must both be considered
in finding the most-probable packing fraction, which is obtained by
minimizing the free energy with respect to $\eta$ (see
equations~(\ref{minfreeA}) and~(\ref{etastq}).

We still must find the dependence of loop length $\ell$ on the amount
of native structure present.
We proceed
by making several approximations for the quantities in equation~(\ref{totnumberstates}).
The result is not sensitive to the exact
values of these quantities.
We approximate the product over loop
lengths in
equation~(\ref{totnumberstates}) by taking a saddle-point value for $\ell$,
effectively
letting all loops have the typical loop
length $\lbar (Q)$. Then $n(\ell|Q) = \delta(\ell - \lbar(Q))
\nL(Q)$ where $\nL(Q)$ is the total number of loops at $Q$. The
typical loop length $\lbar(Q)$ is obtained from the total number of
loops and the total number of disordered residues.
We estimate the total number of disordered residues as a linear
function of $Q$:  $N (1-Q)$. This is a mean-field approximation. In
capillarity models, the deviations from linearity scale as
$N^{2/3}$, but are of order unity for a typical size protein (see
Appendix~\ref{sect:nofQ}).
We estimate the typical loop length $\lbar(Q)$ as the total number of
disordered residues divided by the total number of loops:
\be
\lbar(Q) \cong \frac{N(1-Q)}{\nL(Q)} \: .
\label{eqlstar}
\ee

Generically for small native cores, the number of loops dressing the
native core is proportional to the surface area of the core, which
goes as the number of native residues $N Q$ to the $2/3$ power. However for
large native cores (a nearly folded protein), the unfolding nucleus consists of
disordered protein, so that the number of constraints on loops within
the core (the surface entropy cost) is
proportional to the number of non-native residues $N(1-Q)$ to the $2/3$
power~\cite{PlotkinSS02:quartrev1}. We linearly interpolate between
these two regimes to obtain
\bea
\nL(Q) &\approx& (1-Q) \left[ N Q\right]^{2/3} + Q \left[ N
  (1-Q)\right]^{2/3} +1 \nonumber \\
&\approx& N^{2/3} \left[ Q (1-Q) \right]^{2/3} \left\{Q^{1/3} +
(1-Q)^{1/3}  \right\} +1 \nonumber \\
&\approx& N^{2/3} \left[ Q (1-Q) \right]^{2/3} +1
\label{nL}
\eea
where the expression in curly brackets is approximated as unity since
it varies
between $1$ and about $1.6$ over the range $0\leq Q\leq 1$. One loop
must always be present so that $\lbar(Q)$ remains non-divergent, so
we have explicitly added unity in
equation~(\ref{nL}).
Equations~(\ref{eqlstar}) and~(\ref{nL}) together give
the typical disordered loop length at $Q$ in the model.
Equation~(\ref{nL}) is consistent with previous statements that the
number of loops dressing the folding nucleus scales as
$N^{2/3}$~\cite{FinkelsteinAV97}, however here the $Q$-dependence is
made explicit. When $Q=0$ or $Q=1$, $\nL=1$, and by~(\ref{eqlstar})
$\lbar(0) = N$, and $\lbar(1) = 0$, so the limits behave sensibly.

The entropy of the disordered polymer at $Q$, $S(\eta,Q)$, is then given
by $\nL(Q) S(\eta,\lbar(Q))$, or using equations~(\ref{Scollapse1}),
(\ref{eqsceta}), and~(\ref{eqlstar}),
\bea
S_c (Q,\eta) &=& N(1-Q) \left\{
\ln \frac{\nu}{\eps} - \left(
\frac{1-\eta}{\eta}\right) \ln \left( 1-\eta \right)
- \frac{1}{6} \left[
\left(\frac{\etabar (Q)}{\eta}\right)^{2/3} - 1\right]^2
\right\} \nonumber \\
&\equiv& N (1-Q) \, s_{nn}(Q,\eta)
\label{eq:SQeta}
\eea
where $\etabar(Q) = \lbar(Q)^{-1/2}=[\nL(Q)/N(1-Q)]^{1/2}$.
In equation~(\ref{eq:SQeta}) the
quantity in curly brackets is the entropy per residue for the remaining
disordered polymer at
$Q$. Equation~(\ref{eq:SQeta}) scales extensively with chain length,
which is a consequence of the mean-field approximation made above.

\section{Calculation of the probability distribution of anomalous variance}
\label{sect:pln}

We again write the probability $p(\l,N)$  as a functional integral over
all possible probability distributions, of the probability of a given
distribution $P[ p(x) ]$, times a delta function which counts only
those walks that have a given variance of $\l N a^2$:
\be
p(\l,N) = \int {\cal D} p(x) \: P[p(x) ] \:  \d \left( \l  -
\frac{1}{N a^2}\int \! dx \: x^2 p(x)\right) \: .
\label{functn}
\ee

To obtain $P[p(x) ]$ we imagine dividing the $x$-axis up into bins of
width $dx$, where each bin is labeled by $i$, has coordinate $x_i = i
dx$, and we let $\pbar(x_i) dx \equiv \pbar_i$. The probability after
$N$ trials or events, of a
distribution of numbers $\{ n_i \}$ across all the bins is a
multinomial distribution of essentially infinitely many variables
\be
p\{n_i\} = \frac{N!}{\ldots n_1! \, n_2! \ldots} \cdots \pbar_1^{n_1}
\, \pbar_2^{n_2} \cdots
\ee

Expanding the log of $p\{n_i\}$ to second order, subject to the
constraint that $\sum n_i = N$, and using Stirling's formula, gives
\be
p\{n_i\} = \left(\prod_i 2 \pi N \pbar_i (1-\pbar_i)\right)^{-1/2} \:
\exp\left( - \sum_i \frac{\left(n_i- N \pbar_i\right)^2}{2 N \pbar_i
(1-\pbar_i)}\right)
\ee

This is the distribution in the limit of large N. We apply it with the
understanding that when $N$ is not so large the distribution is an
approximate solution. The approximation is best where $n_i$ is the
largest, which is where the distribution is most appreciable.

In the continuum limit $p\{n_i\} \rightarrow P[p(x)]$, so that
equation~(\ref{functn}) can be written as
\be
p(\l , N) = \frac{1}{2\pi} \int \! dk \, \e^{-i k \l} \int {\cal D} p(x) \:
\e^{\int \! dx \, {\cal L}(p, x, k)}
\ee
where we have Fourier transformed the delta function. The
effective Lagrangian here is
\be
{\cal L}(p, x, k) =
 - N \frac{\left(p(x)-\pbar(x)\right)^2}{2
\pbar(x)} + i k \frac{x^2}{N a^2} p(x)
\ee
where we have used the
fact that the
probability to be within a given slice of width $dx$ is small.

The functional integral amounts to finding the extremum of the
effective action
in the exponent. The extremal probability $p^{\ast}(x) = \pbar(x) + i
k \frac{x^2}{N^2 a^2} \pbar(x)$
and the extremal action $S^{\ast}(k) = \int \! dx \: {\cal
L}(p^{\ast}, x, k) =  - \frac{3}{2 N} k^2  + i k$. The integral over $k$
is then a simple Gaussian integral, so the result for the probability
of anomalous variance is
\be
p(\l , N) = \sqrt{\frac{N}{6\pi}}\: \e^{- N\left(\l -1\right)^2/6} \:
\label{plnapp}
\ee

For a walk in three-dimensions, there are three parameters
characterizing anomalous variance in $x$, $y$, and $z$. Since e.g. steps in $y$ are
uncorrelated from those in $x$, the probability of finding
parameters $\l_x$, $\l_y$, and $\l_z$ is the product of three terms
each of the form~(\ref{plnapp}), but formally with $1/3$ the number of
steps in each of the three dimensions:
\bea
p(\l_x, \l_y,\l_z , N) &=& p(\lx,N/3) \, p(\ly,N/3) \,
p(\lz,N/3)  \nonumber \\
&=& \left(\frac{N}{18 \pi}\right)^{3/2} \: \mbox{e}^{-\frac{N}{18}\left[
    (\lx-1)^2 + (\ly-1)^2 + (\lz-1)^2 \right]}
\label{pln3}
\eea
The variance $\Delta {\bf R}^2$ is given by
\bea
\Delta {\bf R}^2 &=& \Delta x^2 + \Delta y^2 +\Delta z^2 \nonumber \\
&=& \frac{N a^2}{3} \left( \lx + \ly + \lz\right) \nonumber \\
&\equiv& \l N a^2
\eea
so that we seek the probability distribution $p(\l,N)$ of $\l = (\lx + \ly +
\lz)/3$. This is given by
\bea
p(\l ,N) &=& \int d\lx \, d\ly \, d\lz \; \left(\frac{N}{18
  \pi}\right)^{3/2} \: \mbox{e}^{-\frac{N}{18}\left[
    (\lx-1)^2 + (\ly-1)^2 + (\lz-1)^2 \right]} \: \cdot \:
  \d \left( \frac{\lx + \ly + \lz}{3} - \l\right) \nonumber \\
&=& \int d\lx \, d\ly \; 3 \left(\frac{N}{18
  \pi}\right)^{3/2} \: \mbox{e}^{-\frac{N}{18}\left[
    (\lx-1)^2 + (\ly-1)^2 + (3 \l - \lx -\ly -1)^2 \right]} \nonumber
  \\
&=& \sqrt{\frac{N}{6 \pi}} \: \e^{- N\left(\l -1\right)^2/6}
\eea
as in the one-dimensional case.

\section{Number
of disordered residues for a given number of native contacts}
\label{sect:nofQ}

We wish to find the number of disordered residues when a fraction $Q$
of native contacts are present. Equivalently we can find the number of
ordered (native) residues. In the capillarity model this is the
number of residues $\Nnuc$ in the nucleus. The number of native interactions
at $Q$ can be written as the total number of residues $N$ times the mean
number of interactions per residue in the native structure $\zn$,
times the fraction of possible native interactions $Q$.
The number of native interactions
in a capillarity nucleus is the number of interactions in a fully collapsed
(Hamiltonian) walk~\cite{PlotkinSS02:quartrev1}, which has bulk and
surface contributions, giving the equation
\be
N \zn Q = \zb \left( \Nnuc - \s \Nnuc^{2/3} \right) \: ,
\label{nuc1}
\ee
where $\zb$ is the number of native interactions per residue in a
nucleus of infinite size, and $\s$ is the mean fraction of the $\zb$
interactions lost at the surface. In the absence of roughening $\s$
is a very weak function of $N$ and is of order unity. For walks on a
3-D cubic lattice $\s = 1.5$.

In our problem we know the number of native interactions, $N \zn$. We
can find $\zb$ by solving~(\ref{nuc1}) when $\Nnuc =N$:
\be
\zb = \frac{\zn}{1-\s N^{-1/3}} \: .
\label{zb}
\ee
The number of native residues $\Nnuc$ in a capillarity model as a function of
$Q$ is then given by the solution of
\be
\Nnuc - \s \Nnuc^{2/3} = \left( N - \s N^{2/3} \right) Q  \: .
\label{nuccubic}
\ee
Equation~(\ref{nuccubic}) is a cubic equation in $\Nnuc^{1/3}$, with
solution of the form
\be
\Nnuc (Q) = \left[\frac{1}{3} \left( \s + \frac{\s^2}{A^{1/3}} +
  A^{1/3}\right) \right]^3
\label{cubic1}
\ee
where
\bea
A &=& ( B + \sqrt{B^2 - 4 \s^6})/2 \nonumber \\
B &=& 2 \s^3 + 27 N Q - 27 N^{2/3} Q \, \s \nonumber \: .
\eea
Along with the average loop length, the total number of disordered
residues determines the number of loops
at $Q$. A plot of the total number of disordered residues for both the
capillarity model and the linear approximation is shown in
figure~\ref{fignLQ}. One can see from the figure that a linear
approximation for the number of disordered residues is a good one.

\section{Simulation Model and Method}
\label{sect:app-sim}

We introduce non-native interactions to an otherwise energetically
unfrustrated $C_{\alpha}$ model of SH3 domain of {\it src
tyrosine--protein kinease} (src-SH3).  The energetically unfrustrated
model is obtained by applying a G\=o-like Hamiltonian
\cite{Ueda75} to an off--lattice minimalist
representation of the src-SH3 native structure (pdb-code 1fmk, segment
84-140).
We have previously shown that this topology-based model is
able to correctly reproduce the folding mechanism of small, fast-folding
proteins \cite{Clementi2000:PNAS,Clementi2000:JMB}.
A standard G\=o--like Hamiltonian takes into account
only native  interactions, and each of these interactions contributes
to
the energy with the same weight. Protein residues
are represented as single beads centered in their C--$\alpha$
positions. Adjacent beads  are strung together into a polymer chain by
means of bond and angle interactions. The geometry of the native
state is encoded  in the dihedral angle potential and a non--local
potential. The G\=o-like energy of a protein in a configuration $\Gamma$
(with native state $\Gamma_{N}$) is given by the expression:
%
\bea
E(\Gamma, \Gamma_{N})_{\text{G\=o}} 
& = & \sum_{bonds} K_r \left( r - r_N \right)^2 +
\sum_{angles} K_{\theta} \left( \theta - \theta_N \right)^2  + \\
 & + & \sum_{dihedral} K_{\phi}^{(n)}  \left[1 + \cos \left( n \times (\phi -
\phi_0) \right) \right] + \\
 & + & \sum_{i < j -3} \left\{ \epsilon_1(i,j) \left[ 6 \left(
\frac{\sigma_{ij}}{r_{ij}}\right)^{10}  - 5 \left(
\frac{\sigma_{ij}}{r_{ij}} \right)^{12} \right] + \epsilon_2(i,j)\left(
\frac{\sigma_{ij}}{r_{ij}} \right)^{12} \right\}
\label{go_ham}
\eea
where $r$ and $r_N$ represent the distances
between two subsequent residues in, respectively, the configuration
$\Gamma$ and the native state $\Gamma_{N}$. Analogously, $\theta$
($\theta_N$), and $\phi$ ($\phi_0$), represent the angles formed by
three subsequent residues, and the dihedral angles defined by four
subsequent residues, in the configuration $\Gamma$
($\Gamma_{N}$).  The dihedral potential consists of a sum of two terms
for every four adjacent $C_{\alpha}$ atoms, one with period $n =1$ and
one with $n=3$.  The last term in equation (\ref{go_ham}) contains the
non--local native interactions and a short range repulsive term for
non--native pairs (i.e. $\epsilon_1(i,j) = constant < 0$ and
$\epsilon_2(i,j)=0$ if $i$--$j$ is a native pair, while $\epsilon_1(i,j)
= 0$ and $\epsilon_2(i,j)= constant > 0$ if $i$--$j$ is a non--native
pair). The parameter $\sigma_{ij}$ is taken equal to $i$--$j$ native distance
for native interactions, while $\sigma_{ij}=4$\AA\,  for non-native pairs.
Parameters $K_r$, $K_{\theta}$, $K_{\phi}$, $\epsilon$ weight the relative
strength of each kind of interaction entering in the energy and they are
taken to be $K_r = 100 \epsilon$, $K_{\theta} = 20 \epsilon$,
$K_{\phi}^{(1)} = \epsilon$ and $K_{\phi}^{(3)} = 0.5 \epsilon$.

\begin{figure*}[!h]
\centerline{
\includegraphics[width=0.9\linewidth,clip=]{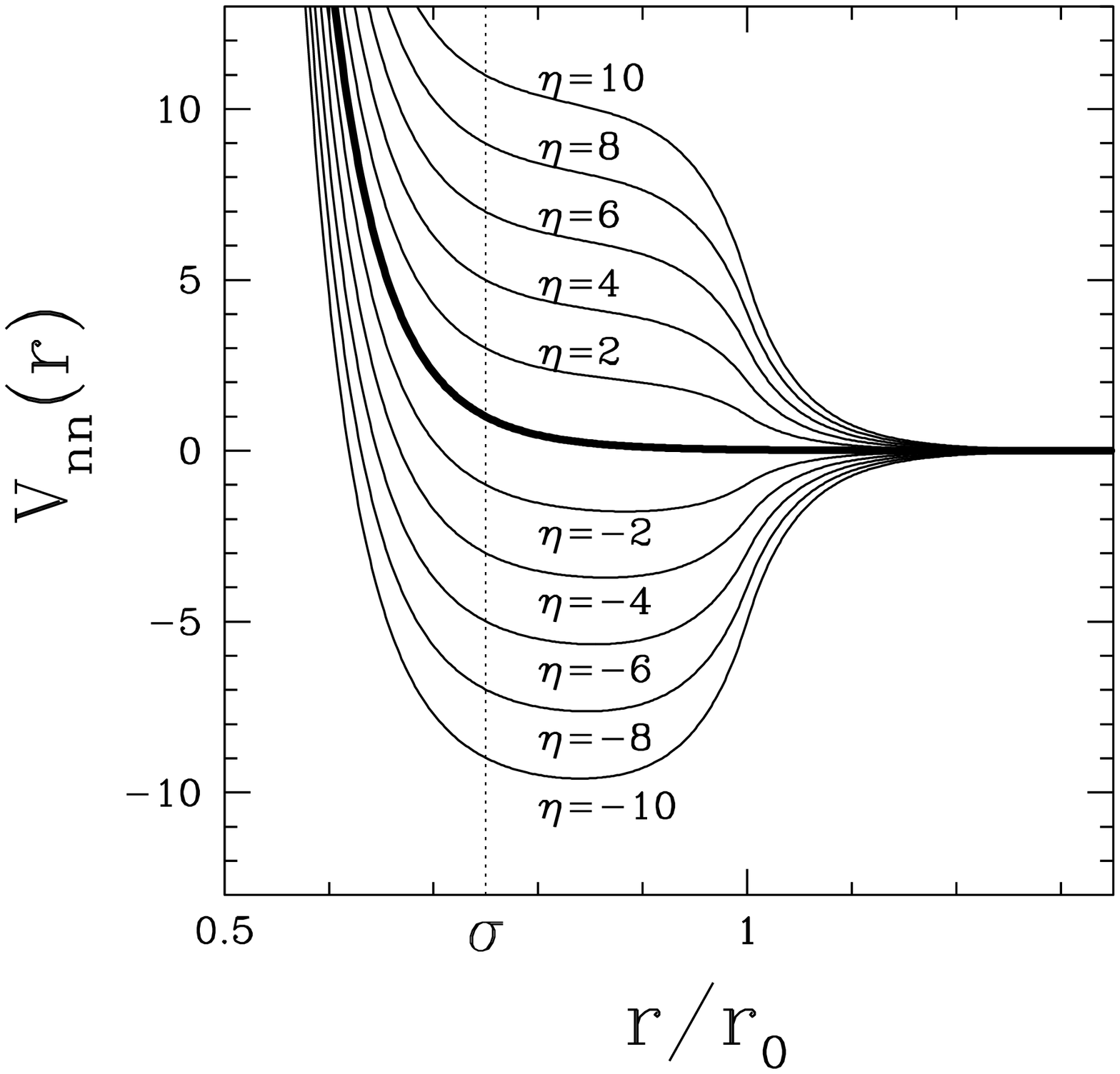}}
\caption{\sf\protect
}
\label{fig_lj}
\end{figure*}

We introduce a progressively increasing perturbation to the
G\=o--like Hamiltonian by replacing the short range repulsive term
in equation (\ref{go_ham}) with attractive or repulsive pairwise
interactions $V_{nn}(r_{i,j})$ in the form:
\begin{equation}
V_{nn}(r_{ij}) =
\begin{cases}
\left(\frac{\sigma_{i,j}}{r_{i,j}}\right)^{12}+
\eta_{\,i,j}\left[1-\frac{1}{2}\left(\frac{r_{i,j}}{r_N}\right)^{20}\right]
& \text{if $r_{i,j}< r_N$},\\
\left(\frac{\sigma_{i,j}}{r_{i,j}}\right)^{12}+
\frac{\eta_{\,i,j}}{2}\left(\frac{r_N}{r_{i,j}} \right)^{20}
& \text{if $r_{i,j}> r_N$}.
\end{cases}
\label{expr_nn}
\end{equation}
Figure~\ref{fig_lj} shows non-native interactions for different values of
the interaction strength $\eta$.
The strength $\eta_{i,j}$ for each non-native pair $(i,j)$ is extracted
randomly from a Gaussian distribution with mean $\epsilon_{NN}$ and
variance $b^2$.
The parameter $\sigma_{i,j}$ in expression \ref{expr_nn} is kept equal to
$4$\AA\, for all non-native interactions, in order to recover the plain G\=o
like Hamiltonian (equation~\ref{go_ham}) in the limit $b\rightarrow 0$,
$\epsilon_{NN}\rightarrow 0$.  The parameter $r_N$ is set to $r_N =
\frac{4}{3}\sigma_{i,j}$. The selected values for $\sigma_{i,j}$ and $r_N$
allow non-native contacts to form in the range of $r_{i,j} \sim 4-5$\AA\,.
The total energy of a configuration $\Gamma$ (with a native state $\Gamma_{N}$),
corresponding to a non-native perturbation strength $b$, is thus:
\be
E(\Gamma, \Gamma_{N})_b = E(\Gamma, \Gamma_{N})_{\text{G\=o}} + \sum_{non-native
(i,j)}  V_{nn}(r_{i,j},\{\eta_b\}),
\label{eq_tot_ham}
\ee
where $\{\eta_b\}$ is a set of quenched variable randomly distributed as
described above.
The case of $b = 0$, $\epsilon_{NN}=0$ corresponds to the unperturbed
G\=o-like representation of  the protein, as it has been studied in
refs. \cite{Clementi2000:PNAS,Clementi2000:JMB}, and we use it as
reference case for comparing the folding rates and folding mechanism.
Sequences with different amount of non-native energy are defined
by progressively increasing the parameter $b$ in the interval  $[0,2]\epsilon$
while keeping $\epsilon_{NN}=0$, or by varying the parameter $\epsilon_{NN}$ in
the interval  $[-1,1]\epsilon$.

The native contact map of a protein is obtained by using the approach
described in ref. \cite{Sobolev96}. Native contacts between pairs of
residues $(i,j)$ with $j \leq i+3$ are discarded from the native map as any
three and four subsequent residues are already interacting in the angle and
dihedral terms.  A contact between two residues $(i,j)$ (native or
non-native) is considered formed if the distance between the $C_{\alpha}$'s
is shorter than $\gamma$ times their equilibrium distance $\sigma_{ij}$
(where $\sigma_{ij}$ = native distance for a native pair,  and
$\sigma_{ij}$ = 4\AA\, for a non-native pair). It has been shown
\cite{Onuchic99}  that the results are not strongly dependent on the
choice made for the cut--off distance $\gamma$.  We have chosen
$\gamma = 1.2$ as in  refs. \cite{Clementi2000:PNAS,Clementi2000:JMB}.
We have used constant temperature Molecular Dynamics (MD) for simulating the
kinetics and thermodynamics of the protein models. We employed the
simulation package AMBER (Version 6) \cite{Amber41:95} and Berendsen
algorithm for coupling the system to an external bath \cite{Berendsen84}.

For each Hamiltonian (obtained for different values of the parameter
$b$), several constant temperature simulations were combined using the WHAM
algorithm \cite{Ferrenberg88,Ferrenberg89}  to generate a specific
heat profile versus temperature and a free energy $F(Q)$ as a function
of the folding reaction coordinates Q and A.
In order to compute folding
rates, several (typically 250) simulations are performed at the
estimated folding temperature for each different sequence.  The
folding time $\tau$ is then defined as the average time interval
between two subsequent unfolding and folding events over this set of
simulations.  The time length of a typical simulation is about
$5\times 10^6$ MD time steps.  In this time range 2 to 5 folding
events are normally observed for the unperturbed G\=o-like protein
model.

The errors (reported as error bars in the plots) on the estimates of
thermodynamic quantities and folding rates are obtained by computing
these quantities from several (more than 100) uncorrelated sets of
simulations and then considering the dispersion of values obtained for
the same quantity.


%
%
\newpage

\vspace{0.5in}
\centerline{TABLES AND TABLE CAPTIONS}
\vspace{1in}
\begin{table}[h]
\begin{tabular}{|c|l|c|c|c|}
\hline
{Symbol} & {\hspace{2.0cm} Meaning} & {Equation$^{\dag}$} & {Simulation values} \\
\hline
$N$ &  Total number of residues in the protein &  (\ref{eqEn}) &  57 \\
$M$ & Total contacts in a fully collapsed globule  & (\ref{eqEn})  & 142\\
$z$ & Average number of contacts per residue & (\ref{eqEn}) & 2.49 \\
$\eps$ & Native energy per contact & (\ref{eqEn})  &  -1.0\\
$\En$ & Energy in the native state & (\ref{eqEn}) &  -142.0\\
$\ln \nu$ & Maximal entropy per residue & (\ref{Scollapse1}) & $\sim$2.4\\
$\enn$ & Mean energy of non-native contacts & (\ref{ne}), (\ref{eq:E}) & 0.0\\
$b^2$ & Energetic variance of non-native contacts & (\ref{ne}),
(\ref{eq:E}) & [0.0 - 4.]\\
$\Tfo$ & G\={o} folding temperature (in energy units) & (\ref{tfo}) & $\sim$1.07\\
\hline
\end{tabular}
\begin{tabular}{lcc}
${}^{\dag}\,$  {\footnotesize{Equation where the symbol is first defined, or representative equation. }}& &\\
\end{tabular}
\label{table1}
\caption{
Table of values for parameters in the model
}
%
\end{table}

\newpage

\vspace{0.5in}
\centerline{FIGURE CAPTIONS}

\vspace{0.6in}

\begin{figure*}[!h]
\centerline{
\includegraphics[width=0.95\linewidth,clip=]{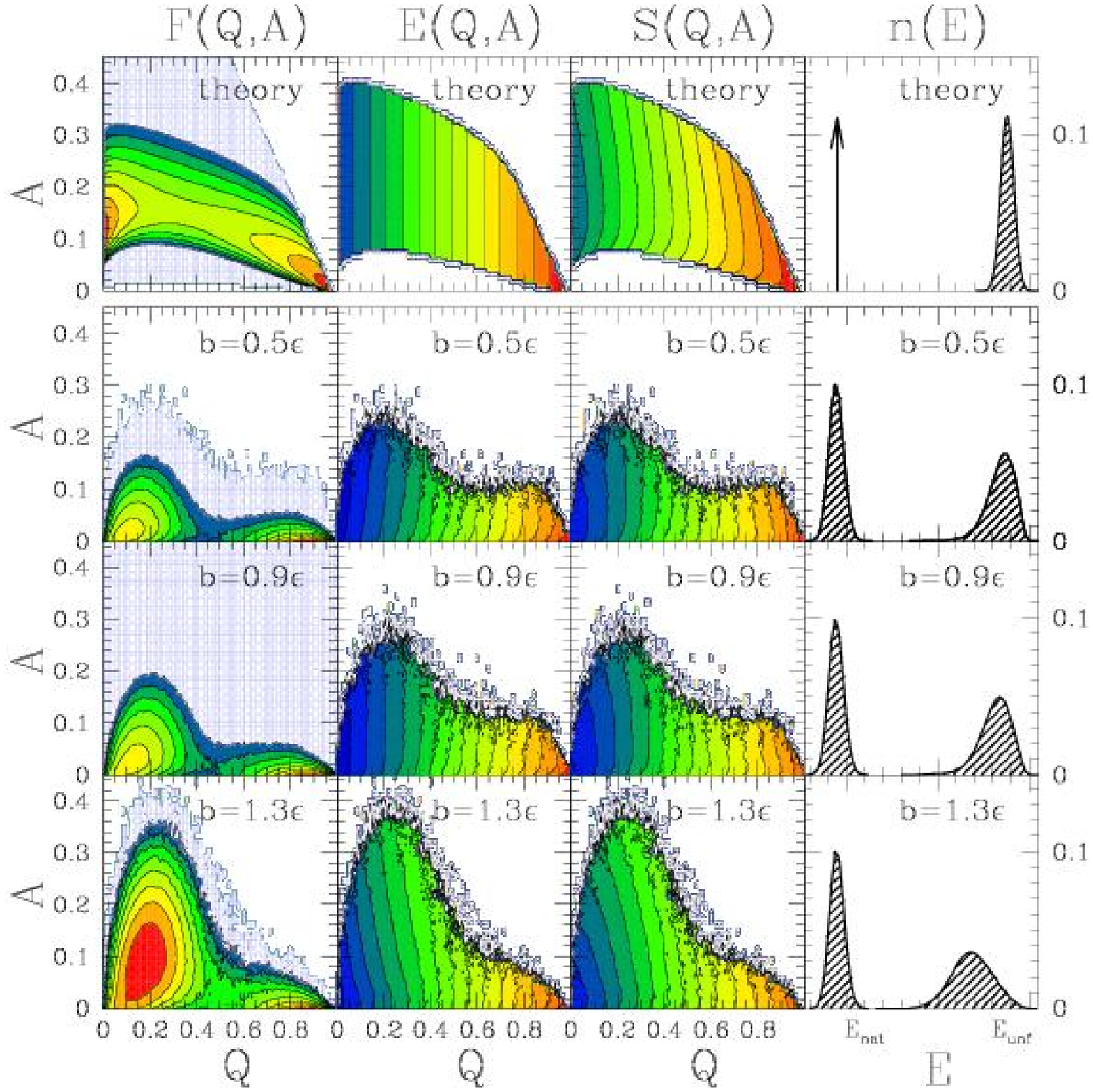}}
\caption{\sf\protect
}
\label{esf}
\end{figure*}

\begin{figure*}[!h]
\centerline{
\includegraphics[height=0.33\linewidth,clip=]{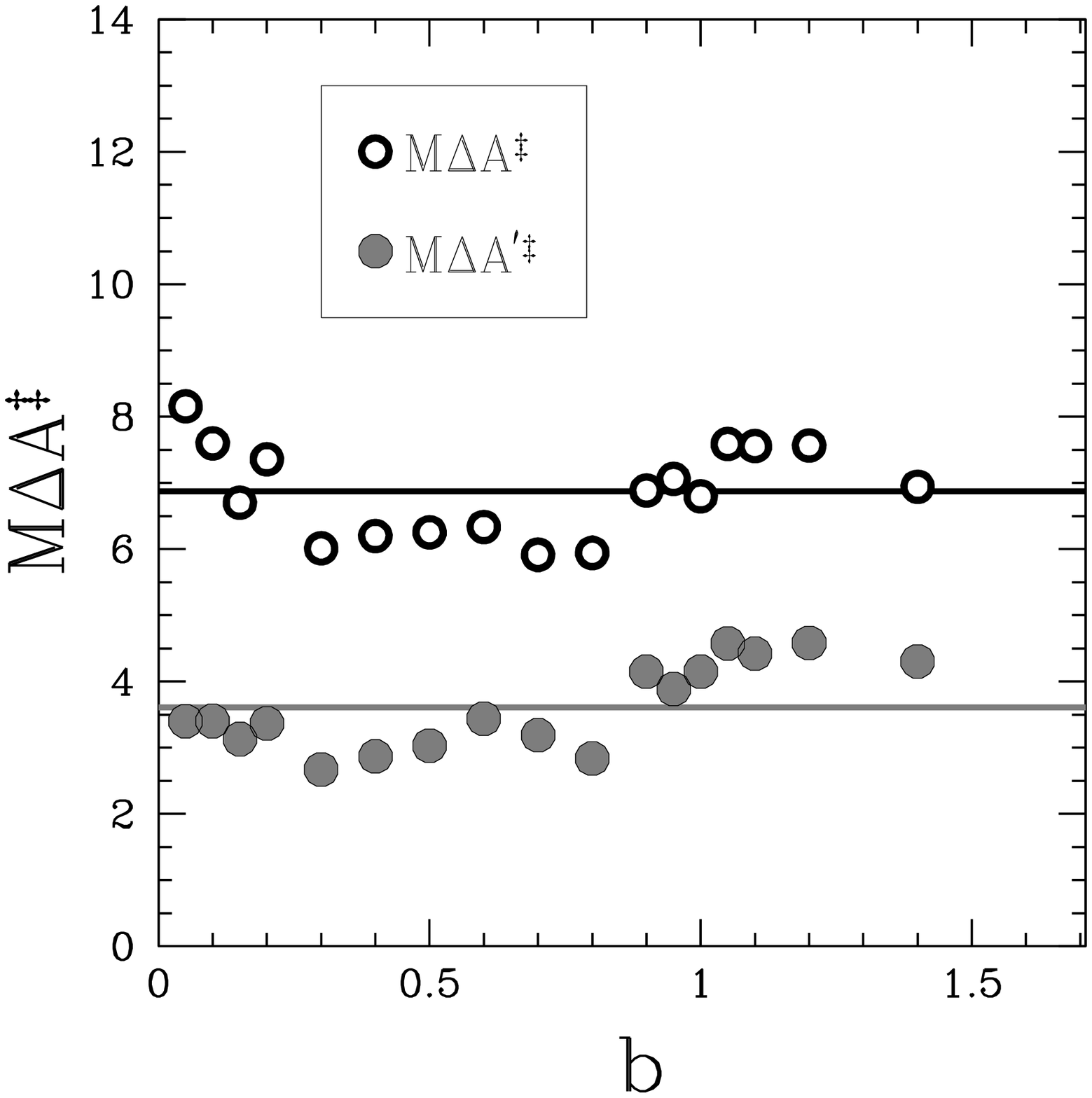}
\includegraphics[height=0.33\linewidth,clip=]{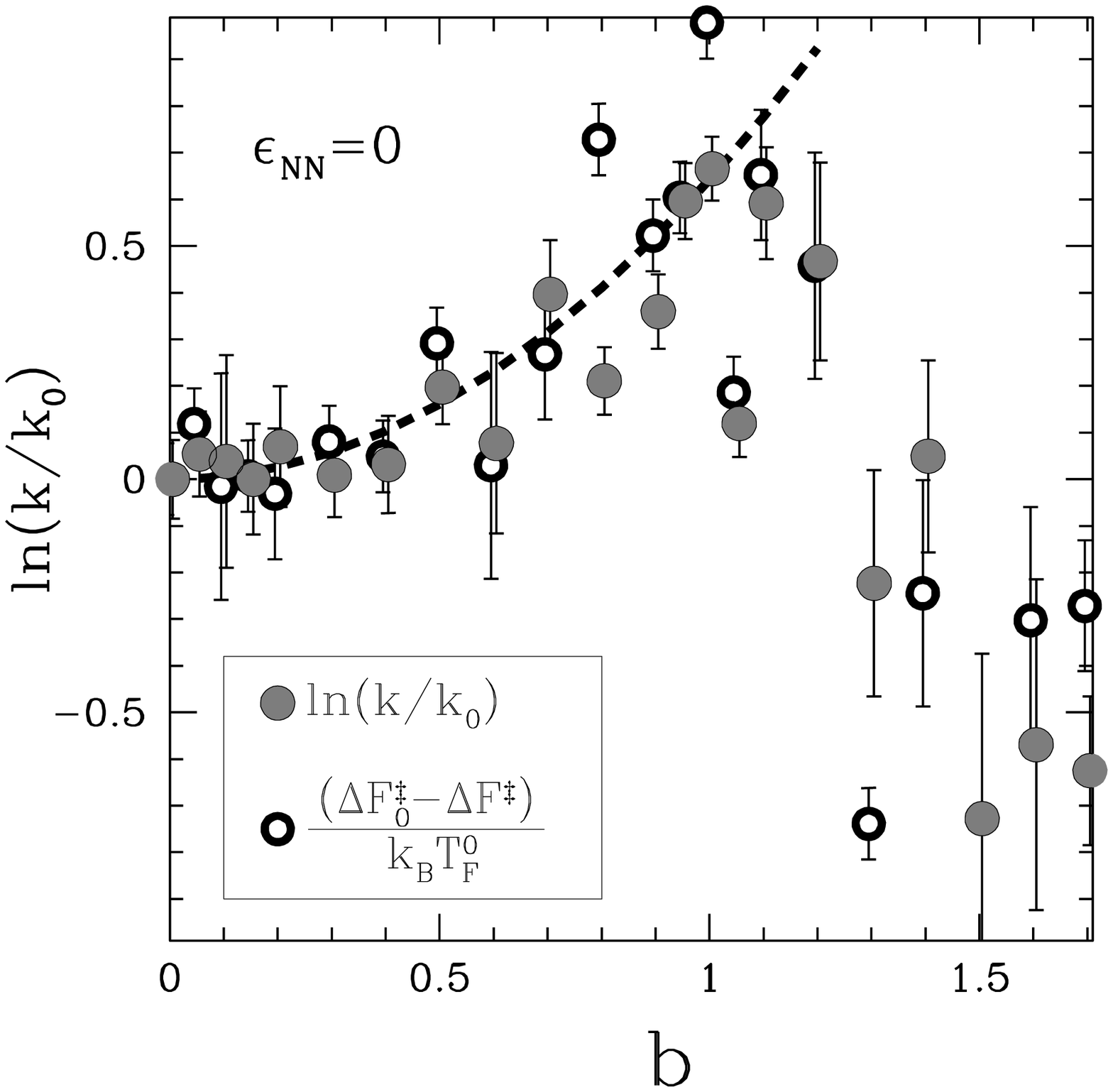}
\includegraphics[height=0.33\linewidth,clip=]{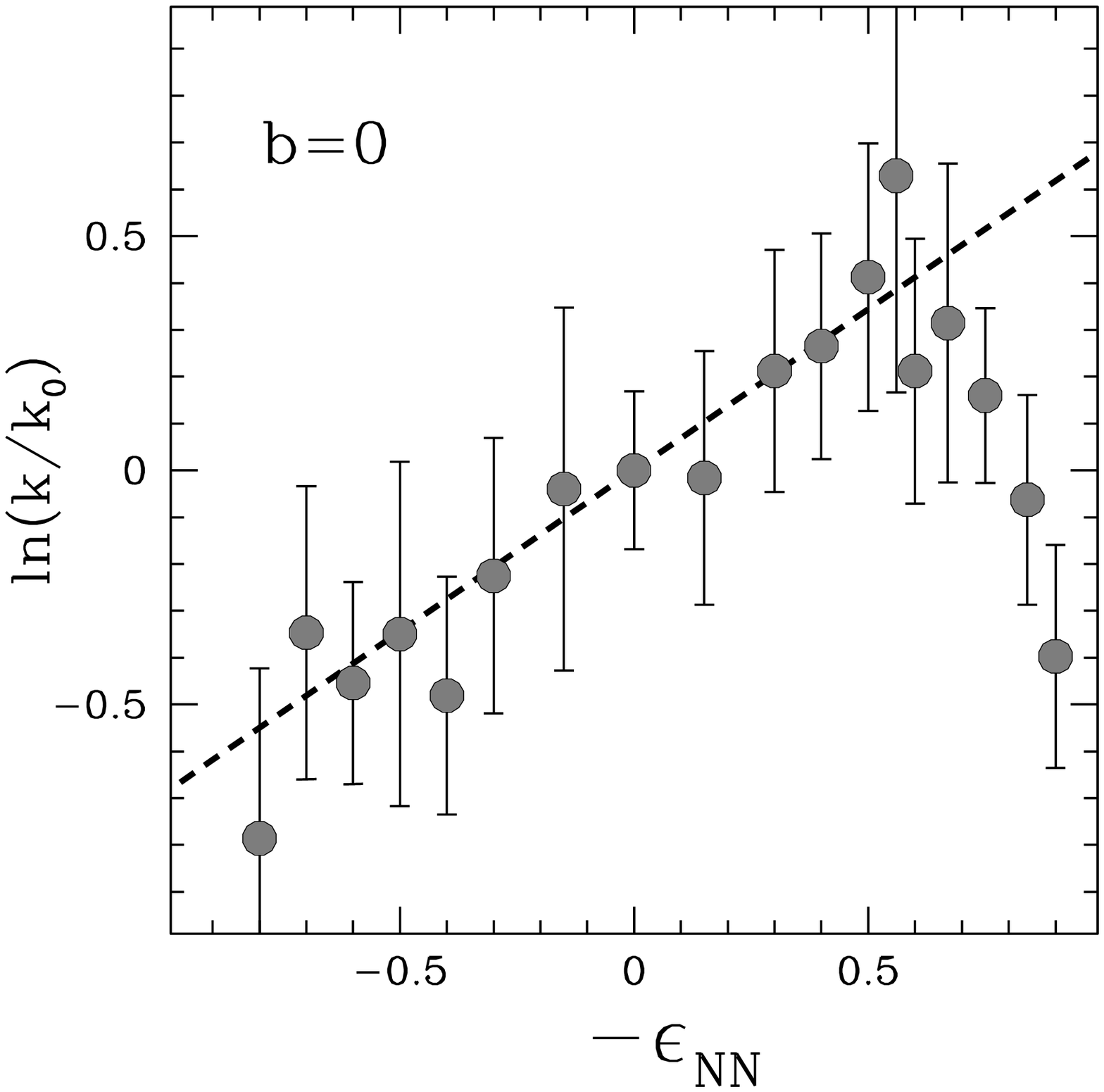}}
\caption{\sf\protect
}
\label{fig_rate}
\end{figure*}

FIGURE \ref{esf}: Free energy (first column from the left), energy
(second column), and entropy (third column) surfaces as functions of
the fraction $Q$ of native contacts, and the fraction $A$ of
non-native contacts, as obtained from theory and simulations. Also
shown is the fraction of states, $n(E)$, populated as a function of
the energy $E$. The distribution of the energy in the unfolded
ensemble is shown, along with the
distribution in the native state (forth
column). The distribution $n(E)$ is normalized, \ie the integral
of $n(E)$ over all energies is 1 in all the cases plotted here.
All free energy contours are spaced at about $1\epsilon$ (where $\epsilon$
is the energy per native contact).
Values of the parameters are given in table I.\\ 
Top row:
Theoretical free energy, energy, and entropy surface at the folding
temperature, obtained from equations (\ref{eq:F}) and
(\ref{eq:SQeta})) with the all parameters set equal to the
corresponding simulation values (see table I) and
$b_{theo}=0.3\epsilon$ where $\epsilon$ is the energy per native
contact (this corresponds to $ 0.9 \epsilon< b < 1.3\epsilon$ in the
simulations, see text for detail). 
The transition state has more non-native contacts than the unfolded
state. The difference in the theoretical model is $\Del A^{\ddag}\simeq 0.035$.
This amounts to an increase in the total number of non-native contacts
of $M A\simeq 5$. The barrier height is about $3.47 \epsilon \simeq 3.3\; \kboltz T_f$.\\
Bottom 3 rows :  corresponding results obtained from simulations,
for three different values of the non-native energy perturbation parameter
$b$: $b=0.5\epsilon$ (second row), $b=0.9\epsilon$ (third row), and
$b=1.3\epsilon$ (bottom row). Barrier heights and values of $\Del A^{\ddag}$ obtained 
in simulations are plotted in figure~\ref{fig_rate} as a function of the non-native
energy perturbation parameter $b$.

\vspace{0.6in}

\begin{figure*}[!h]
\centerline{
\includegraphics[width=0.9\linewidth,clip=]{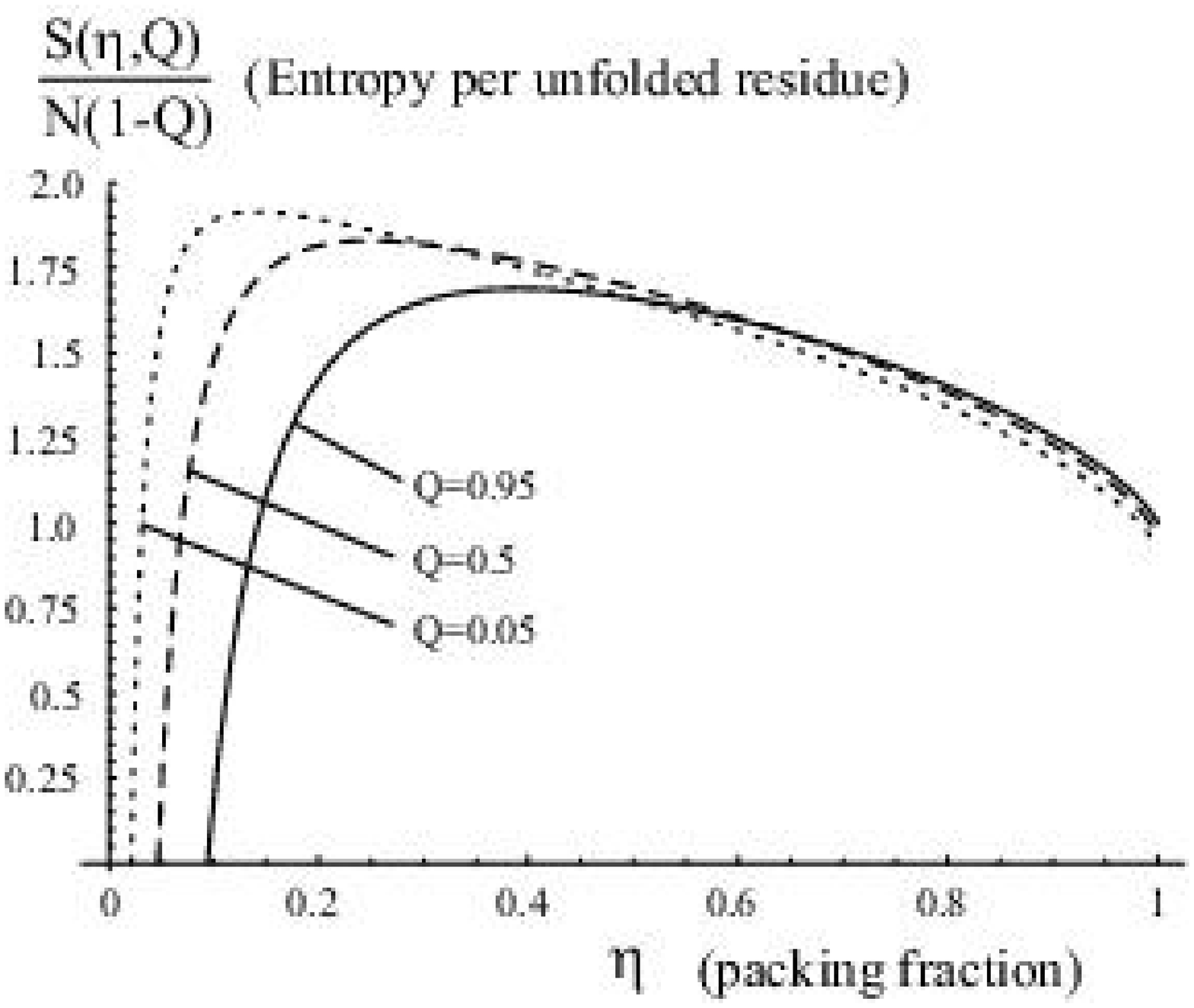}}
\caption{\sf\protect
}
\label{seta}
\end{figure*}

FIGURE \ref{seta}:
The entropy per residue $s_{nn} (Q,\eta) = S_c(q,\eta)/N(1-Q)$ in
equation~(\ref{eq:SQeta}), for the disordered part of a
protein of nativeness $Q$, as a function of the disordered polymer's
packing fraction $\eta$.

\vspace{0.6in}

\begin{figure*}[!h]
\centerline{
\includegraphics[width=0.9\linewidth,clip=]{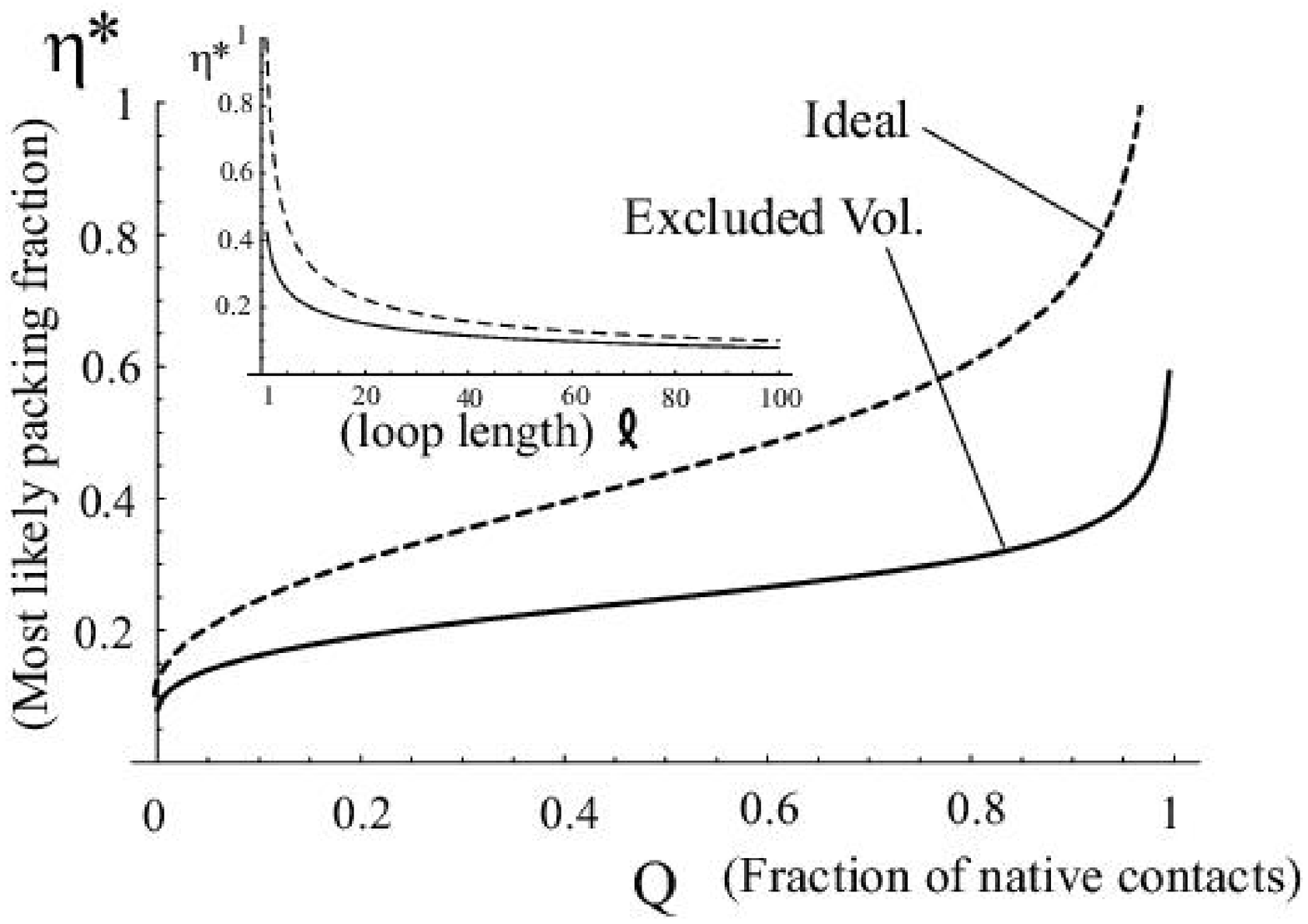}}
\caption{\sf\protect
}
\label{figetaq}
\end{figure*}

FIGURE \ref{figetaq}:
The most probable packing fraction $\etast$ is a monotonically
increasing function of nativeness $Q$. The dashed curve shows the
characteristic packing fraction when the disordered loops are
assumed to obey ideal chain statistics. The solid curve accounts for
the effects of excluded volume, which are included in
equation~(\ref{eq:SQeta_bis}). Inset: The most probable packing fraction
is a decreasing function of the mean disordered loop length $\lbar$ in
equation~(\ref{eqlstar}).

\vspace{0.6in}

\begin{figure*}[!h]
\centerline{
\includegraphics[width=0.9\linewidth,clip=]{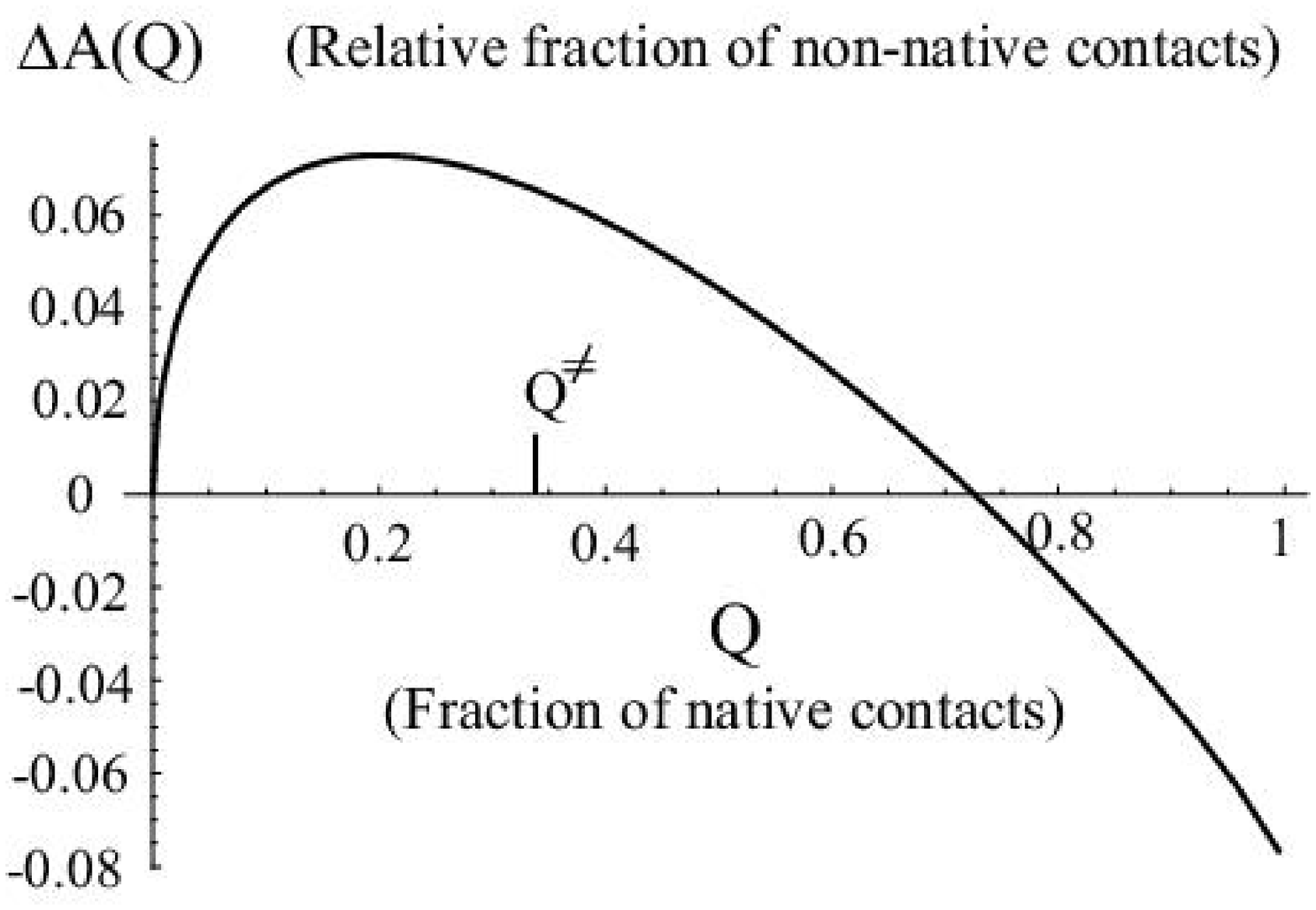}}
\caption{\sf\protect
}
\label{delaq}
\end{figure*}

FIGURE \ref{delaq}:
Fractional change in the number of non-native interactions as
a function of nativeness $Q$, for the theoretical model. We can see that the number of non-native
interactions initially increases before decreasing. For the model
considered here, the barrier
position $Q^{\neq}$ is well within this region of values where the number
of non-native
interactions has increased. There are generically more non-native
interactions present in the transition state than in the unfolded
state, for strongly minimally frustrated proteins. The effect is
fairly modest- for a hundred
residue protein there are about $6$ more non-native interactions in
the transition state. The shape of the curve is obtained from setting
$\left. \D F/\D A\right|_{Q} = 0$ in first row of figure~\ref{esf}. 

\vspace{0.6in}

\begin{figure*}[!h]
\centerline{
\includegraphics[height=0.90\linewidth,clip=]{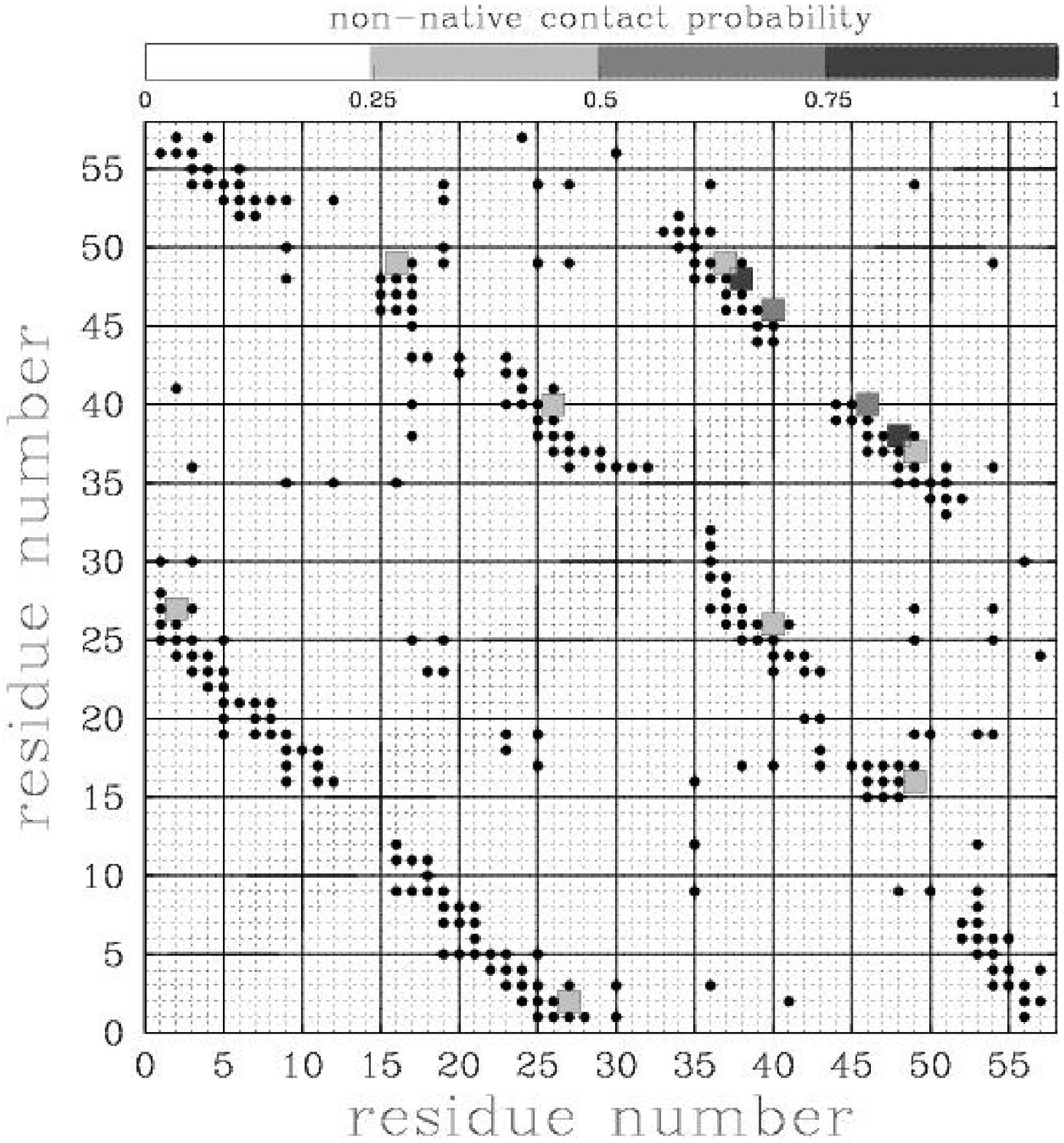}}
\caption{\sf\protect
}
\label{fig_nnformed}
\end{figure*}

\begin{figure*}[!h]
\centerline{
\includegraphics[height=0.49\linewidth,clip=]{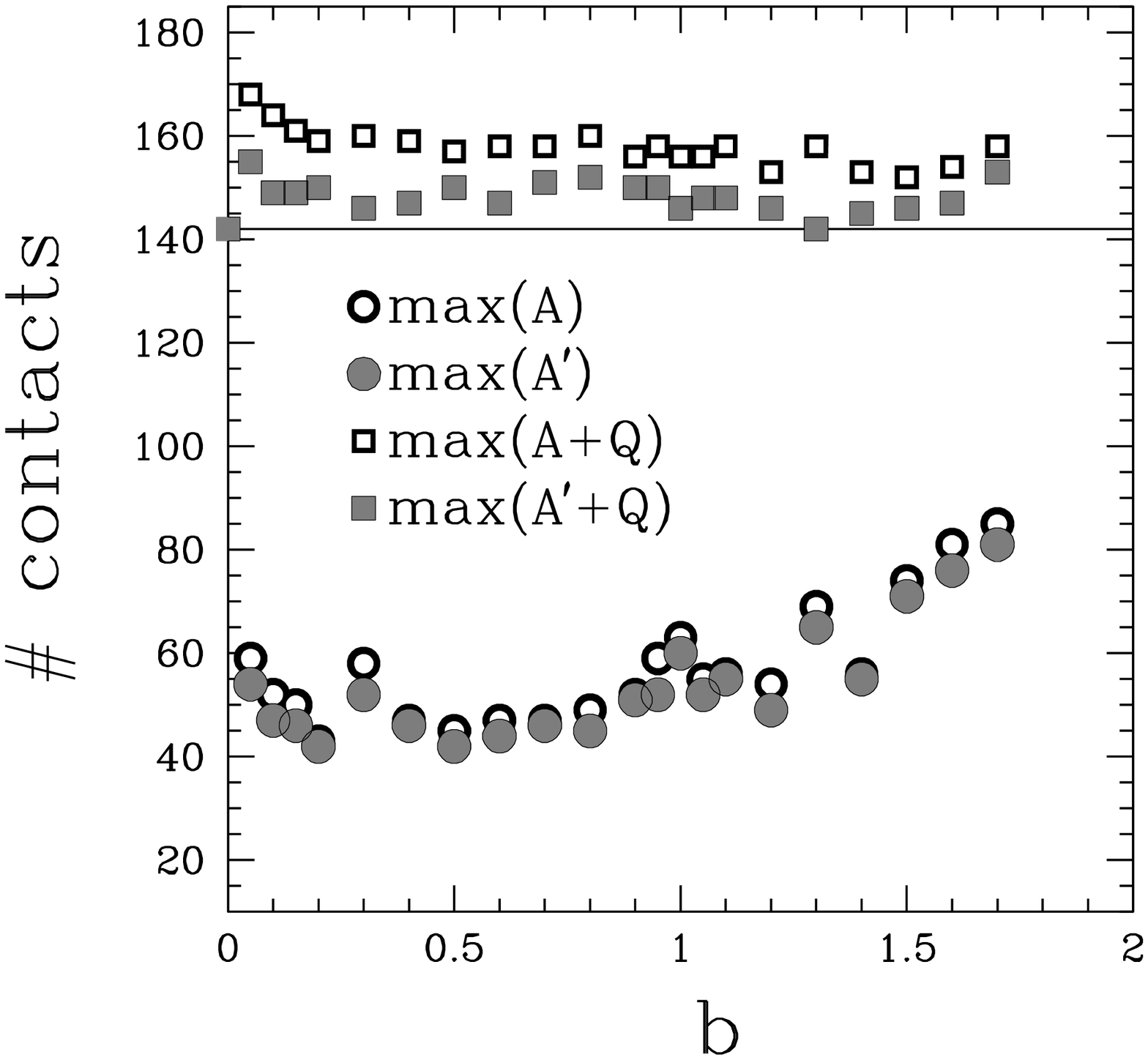}
\includegraphics[height=0.49\linewidth,clip=]{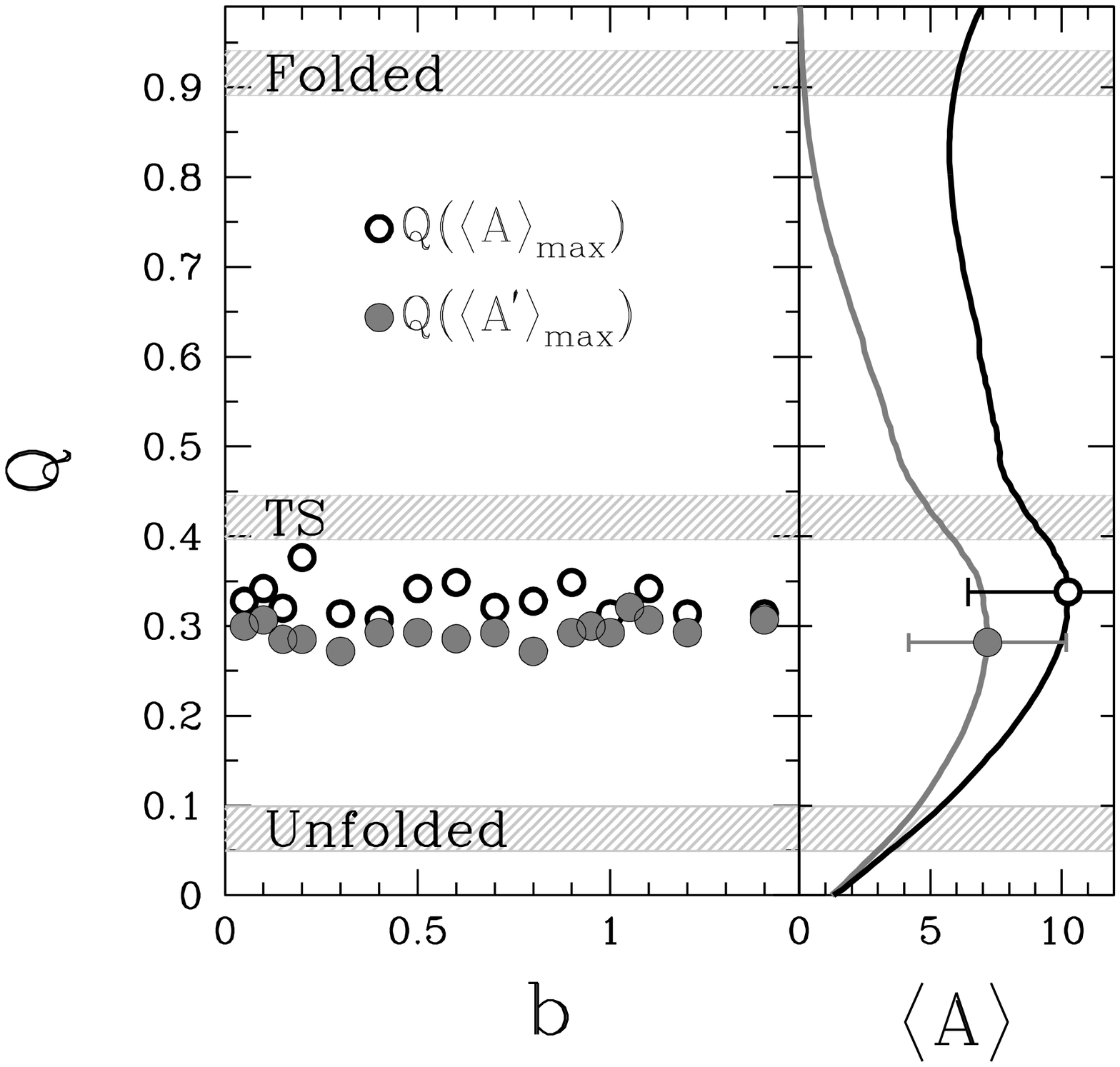}}
\caption{\sf\protect
}
\label{fig_Amax}
\end{figure*}

FIGURE \ref{fig_nnformed}:
Probability of formation of non-native contacts in the native
configuration of SH3.
Black dots in the contact map represent native contacts, non-native contacts
formed with probability higher than 0.25 are color-coded according to the
gray-scale on top.
Probability values are computed by averaging the formation of
non-native contacts over $\gtrsim 50.000$ configurations with $Q > 0.9$
from folding/unfolding simulations. The data shown in this figure are for a
non-native perturbation strength $b/\epsilon=1.3$. Similar results are
obtained for different values of the parameter $b$ (see figure~\protect\ref{fig_Amax}).

\vspace{0.6in}

FIGURE \ref{fig_Amax}:
(a) The lower panel shows the maximum number of non-native contacts registered in simulations
for different values of the perturbation parameter $b$ (in units of $\epsilon$). Black circles
dots indicate
the maximum in the reaction coordinate $A$, while filled gray dots correspond to the corrected
coordinate $A^{\prime}$ (see text for details). The maximum number of all contacts
(both native and non-native) is shown in the upper panel, for different values of $b$.
Empty black squares indicate the maximum value obtained when all contacts separated
by at least three residues
along the sequence are considered (\ie $Q+A$), filled gray squares correspond
to the values obtained when non-native contacts likely to be
formed in the native structure are removed (\ie $Q+A^{\prime}$).\\
(b) Right panel: Average number of non-native contacts $\langle A \rangle$
formed in simulations as a function the number of native contacts formed, for a
perturbation parameter $b/\epsilon = 0.5$. Horizontal bars at the maximum of $\langle A \rangle$
correspond to the standard deviation around the average at the peak value. The black curve correspond
to the coordinate $A$, while the gray line to $A^{\prime}$. The values of $Q$ corresponding to
the maximum of $\langle A \rangle$ and $\langle A^{\prime} \rangle$
are shown in the left panel, for
different values of the perturbation parameter $b$ (black circles and filled gray dots, respectively).

\vspace{0.6in}

\begin{figure*}[!h]
\centerline{
\includegraphics[height=0.45\linewidth,clip=]{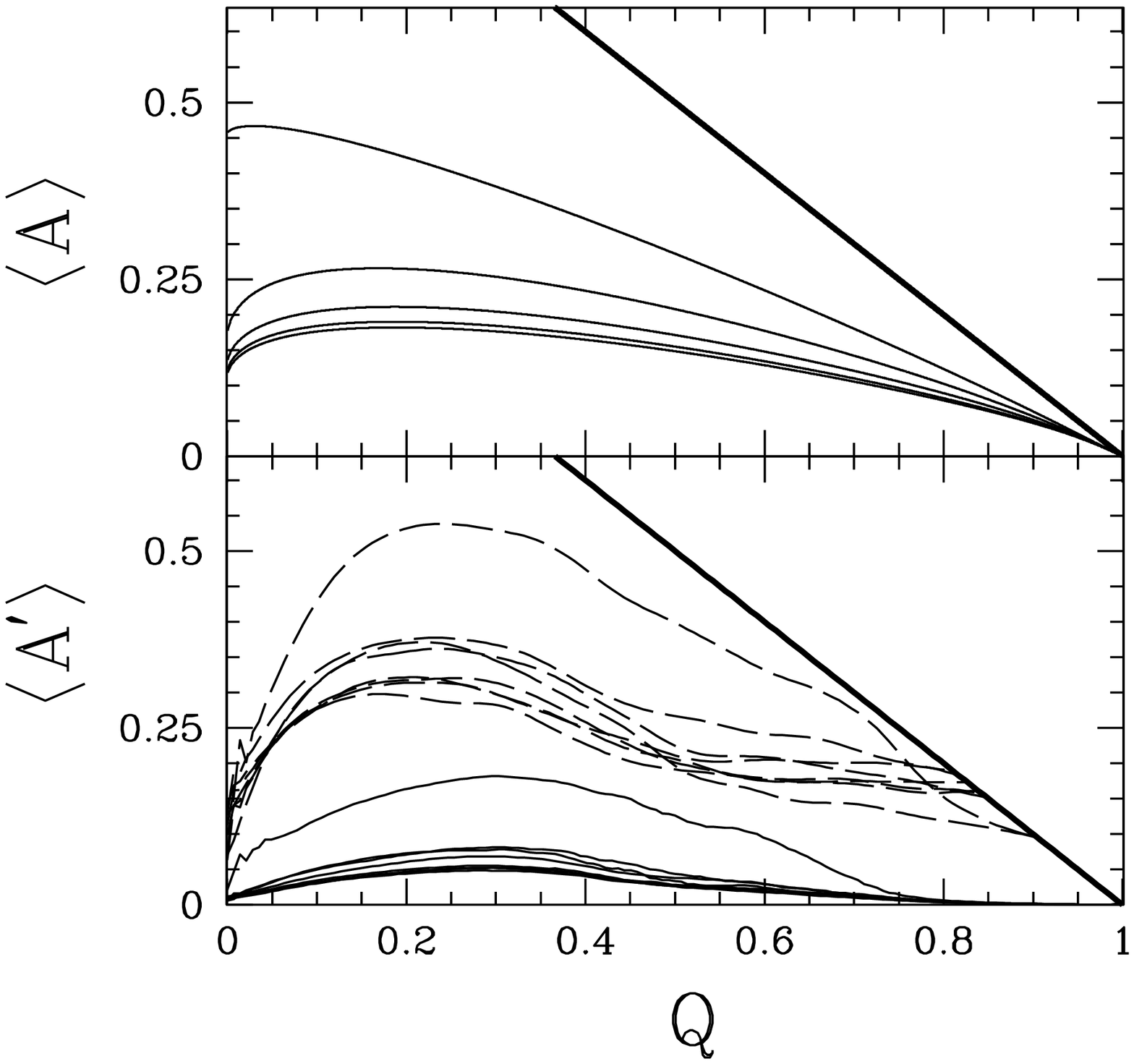}
\includegraphics[height=0.45\linewidth,clip=]{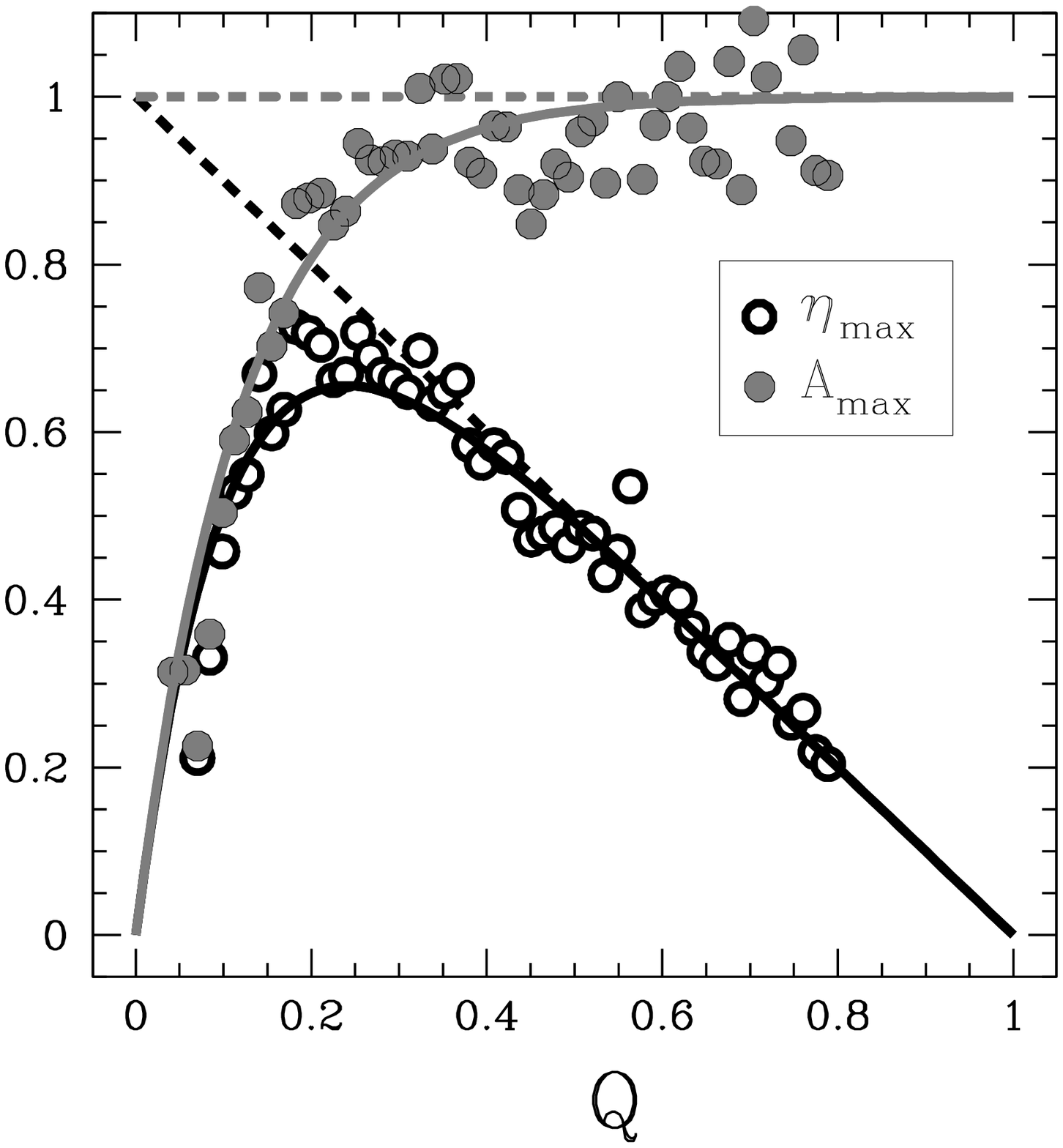}
}
\caption{\sf\protect
}
\label{fig_Asim}
\end{figure*}

FIGURE \ref{fig_Asim}:
(a) Upper panel: Continuous curves illustrate the behavior of $\langle A
\rangle$ vs. Q as predicted by the theory (with all parameters set to the
simulations values, see table I). The different curves correspond to values
of $b/\epsilon=0.1,0.3,0.5,0.7,0.9$ (increasing values of $b$ lead to
higher values of $\langle A\rangle$). The thick black line represents the
maximum value of $A$ allowed in the theory at different values of $Q$
(independent on the value $b$). Lower panel: $\langle A^{\prime} \rangle$
vs. Q (continuous curves) as obtained from simulations, for values of
$b/\epsilon=[0.2,1.6]$ (increasing values of $b$ lead to higher values of
$\langle A^{\prime}\rangle$).  Dotted curves represent the highest values
of $A^{\prime}$ found in simulations at different values of $Q$, for the
same values of $b/\epsilon$.  The maximum value of $A$ allowed in the
theory is also plotted for comparison (thick black line). \\ 
(b) Filled gray dots show the maximum value of $A^{\prime}$ detected in all
equilibrium and quenched simulations for many values of $b$ (see text), as
a function of the fraction of native contact formed, $Q$. Black circles
correspond to the maximum packing fraction of the non-native part of the
protein, as obtained by using equation~(\ref{eqAeta}), \ie $\eta_{max} =
A^{\prime}_{max}/(1-Q)$. Dotted lines show the maximum values for $A$ (in
black) and $\eta$ (in gray) allowed in the theory. Continuous lines in the
corresponding colors represents the best fit of the data to a
phenomenological exponential decay of $A^{\prime}_{max}$ at small values of
$Q$: $A^{\prime}_{max} = (1-Q)[1-\exp{(-Q/Q_{c})]}$. Regression analysis
yields $Q_{c} = 0.12$.  The best fit for $A^{\prime}_{max}$ is shown in
gray, in black for $\eta_{max}$.

\vspace{0.6in}

\begin{figure*}[!h]
\centerline{
\includegraphics[height=0.31\linewidth,clip=0.05]{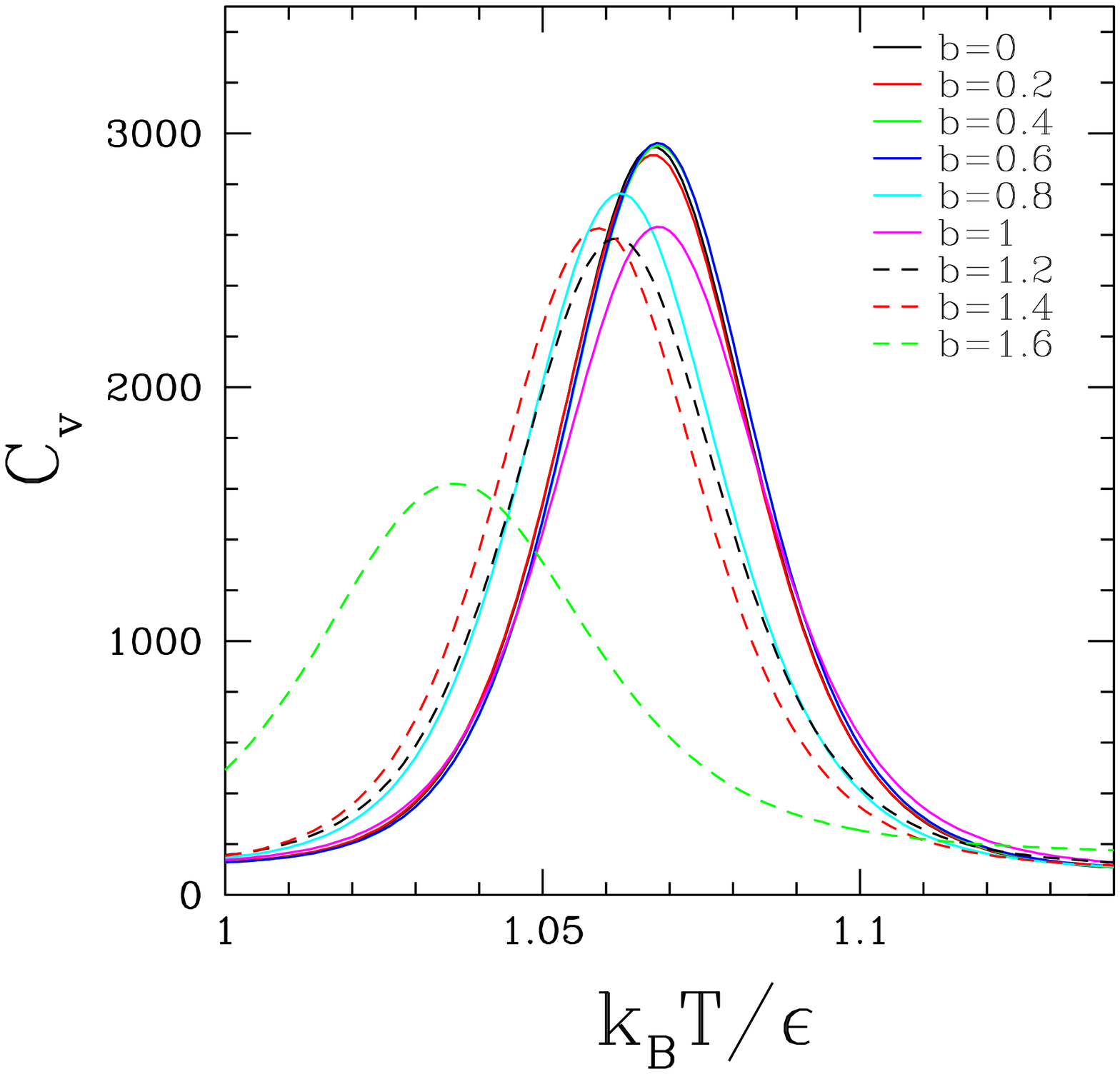}
\includegraphics[height=0.31\linewidth,clip=]{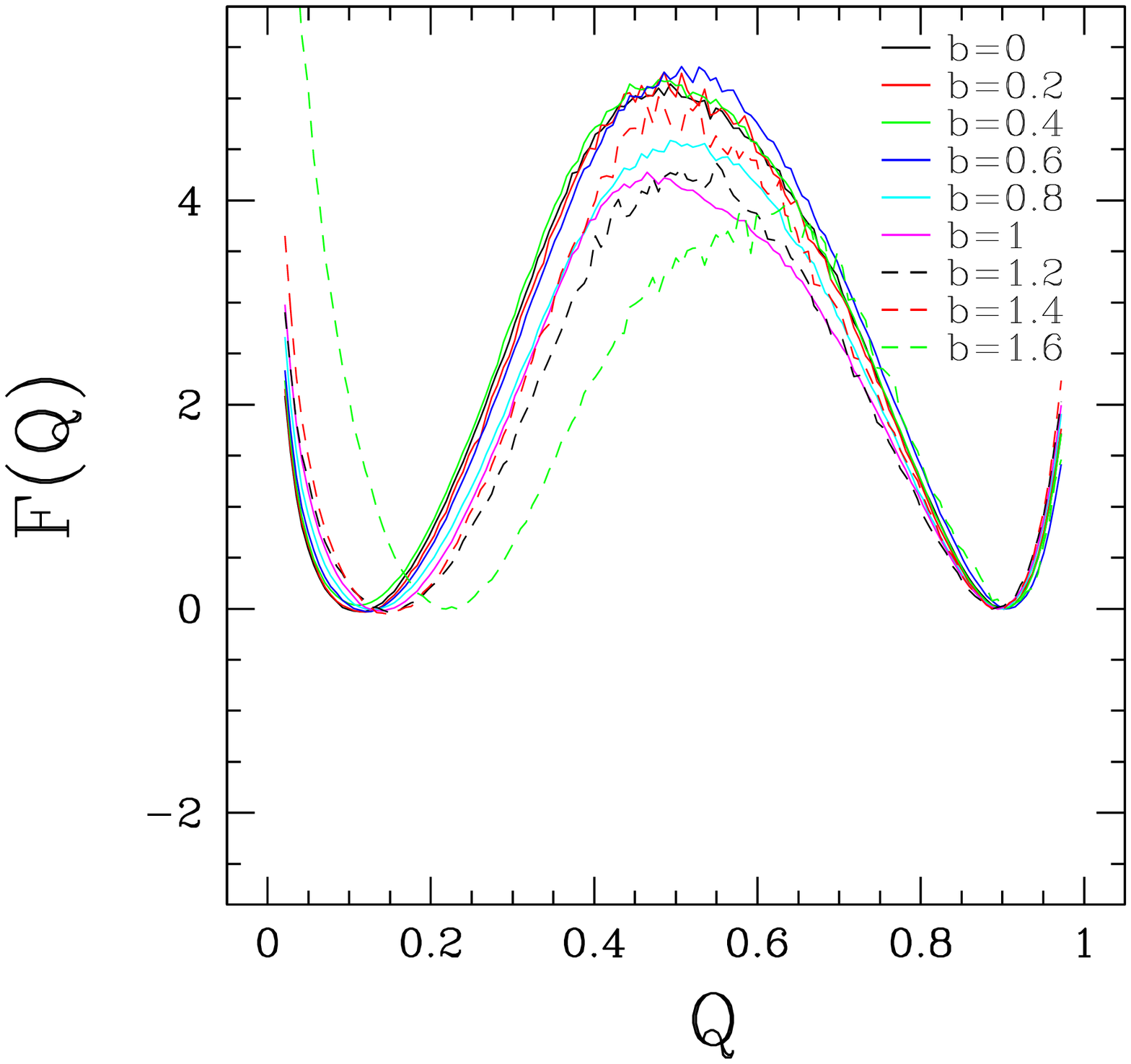}
\includegraphics[height=0.31\linewidth,clip=]{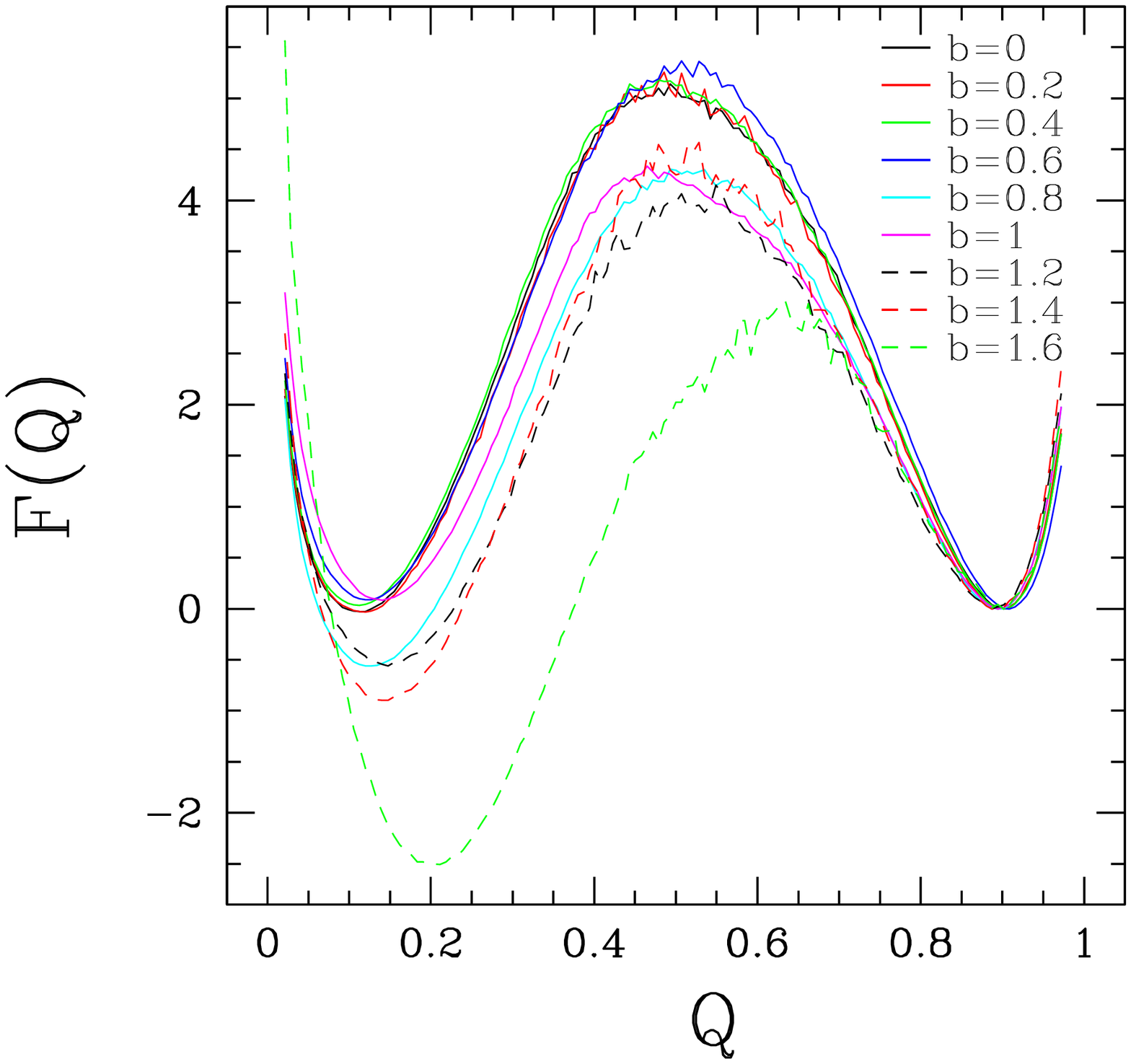}}
\caption{\sf\protect
}
\label{F_vs_Q}
\label{fig:cv}
\end{figure*}

FIGURE \ref{F_vs_Q}:
(a) Heat capacity as a function of temperature, as obtained from simulations
for different values of the parameter $b$. Temperature is measured in units of
native energy per contact, $\epsilon$.
(b) Free energy as a function of the fraction $Q$ of native contacts,
as obtained from simulations for several different values of the non-native
energy perturbation parameter $b$. Free energy curves for all values of $b$
shown in (b) are obtained at their corresponding folding
temperatures $T_f(b)$ (estimated from the heat capacity curves, plotted in (a)),
while all curves in (c) are at the folding temperature of the unperturbed case $T_f^0 = T_f(b=0)$.

\vspace{0.6in}

\begin{figure*}[!h]
\centerline{
\includegraphics[height=0.49\linewidth,clip=]{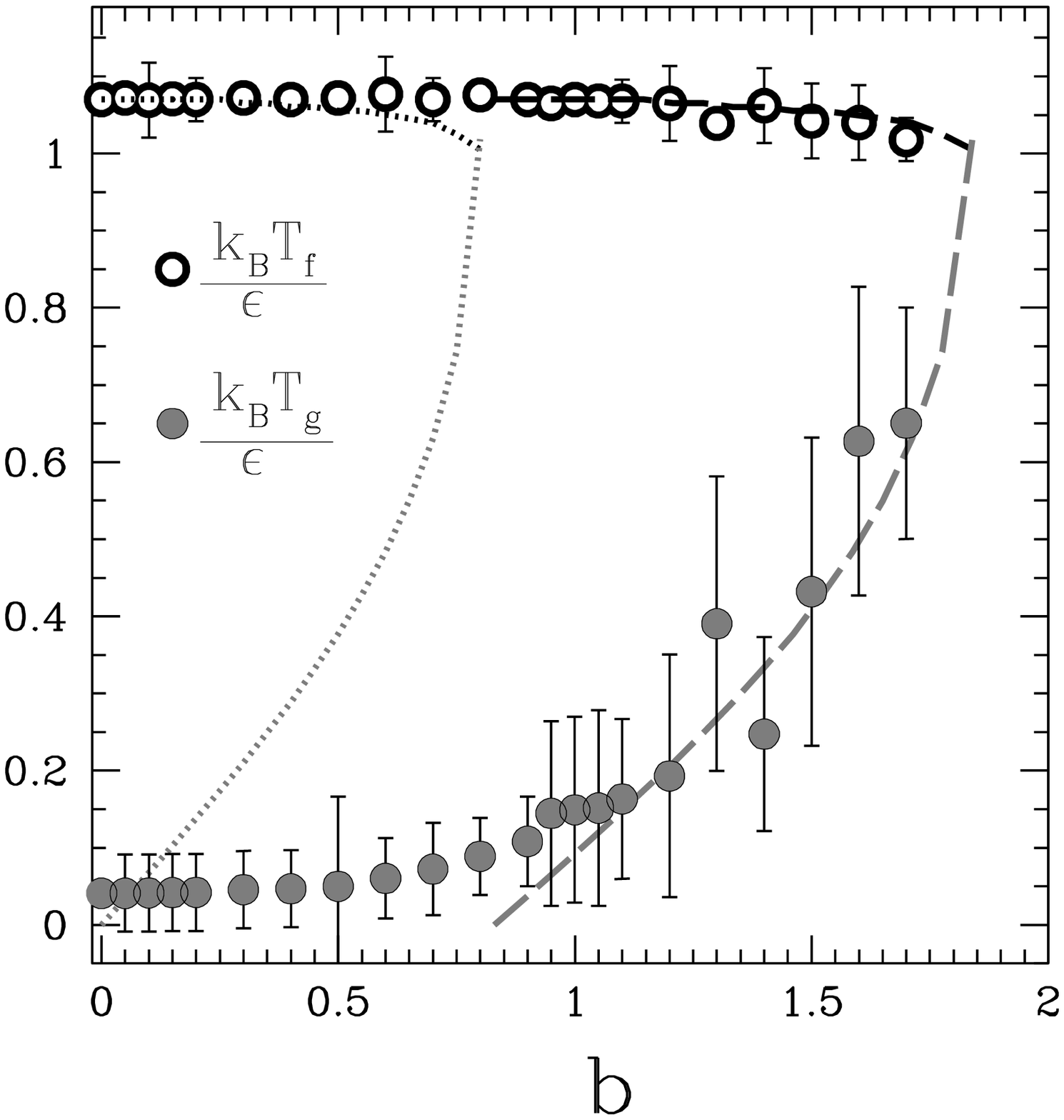}
\includegraphics[height=0.49\linewidth,clip=]{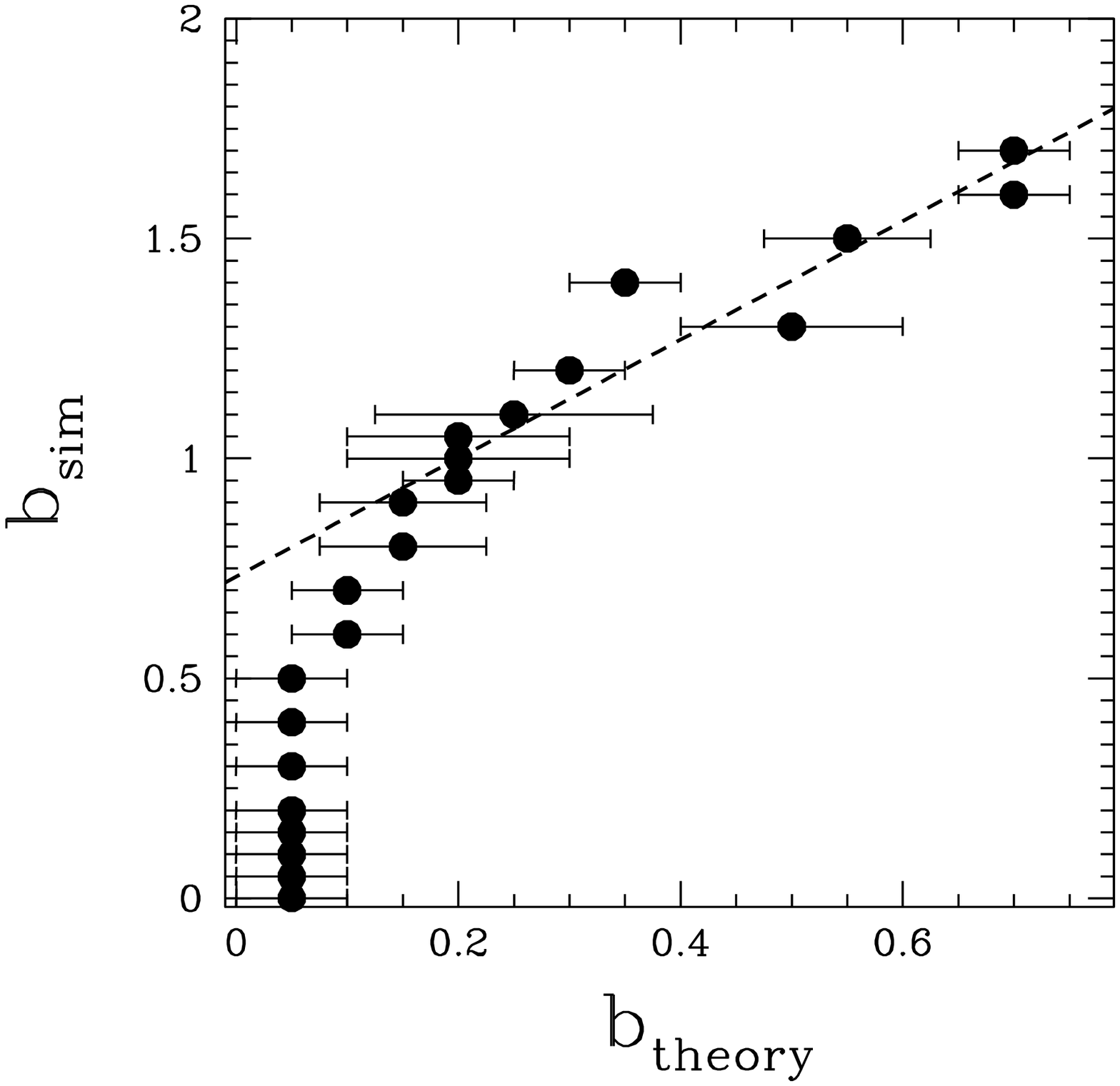}}
\caption{\sf\protect
}
\label{fig_tf}
\label{fig_glass}
\end{figure*}

FIGURE \ref{fig_glass}:
(a) Folding temperature $T_f$ (black circles) and glass temperature $T_g$ (filled gray dots), from
simulation of the perturbed G\=o-model, as a function of the non-native energy perturbation
strength $b$ (with $\epsilon_{NN} = 0$). Dotted lines represent the theoretical
prediction for $T_f$ (black line) and $T_g$ (gray line), when all the parameters of the theory
are set equal to the simulation parameters (see table I). Dashed lines represent the
best fit of the simulation data to the theoretical prediction (see (b)).
Temperatures and energies are measured in units of $\epsilon$, the native
energy per contact.
The folding temperature is almost constant in the range shown,
while the glass temperature raises from zero (plain G\=o model, with
no energetic frustration), to values close to $T_f$ for a high level
of non-native perturbation. As $T_g/T_f$ approaches 1, several
non-native low energy states compete with the native state and the
folding is dramatically slowed down. Moreover, as $T_g/T_f \rightarrow 1$
the system is no longer self-averaging and different realizations
of the non-native perturbation can lead to different results.
There is a wider range where $T_f/T_g >1$ in the simulations than in
the theory, indicating a larger range of $b$ where rate enhancement
effects may be seen.
(b) The same destabilizing effect on the folding predicted in the theory (in terms
of $T_f/T_g$), for a given value of the parameter $b$, is observed in simulations for a much larger
value of $b$. All values of $b$ used in simulation ($b_{sim}$) are plotted in this
figure as a function of the values of $b$ yielding the same glass
temperature in the theory ($b_{theory}$) (see text for details).
The dashed line represents the best fit
of the data to the expression $b_{sim} = \alpha b_{theory} + \beta$, for
$b_{sim} > 0.8$. The result from this fit is also shown in (a).

\vspace{0.6in}

FIGURE \ref{fig_rate}:
(a) Average difference in the number of non-native contacts in the transition state
and unfolded state ensemble, as detected from simulation, as a function of the perturbation
parameter $b/\epsilon$. The values obtained by considering all non-native contacts are shown
as dark circles, while filled gray dots correspond to the corrected values
(\ie considering only non-native
contacts that are not formed in the folded state, see text for details). The average of these
quantities over all the considered values of $b/\epsilon$ are plotted as continuous straight lines
of the same color. These numbers are comparable to the theoretical
estimates of $\sim 6$ more non-native contacts in the transition state
for a $100$-residue protein (see figure~\ref{delaq}).
\newline
(b) Barrier height $\Delta F^{\dag}$ (black circles) and log folding
rate $k$ (filled gray dots) as a
function of the non-native energy perturbation strength $b$ (for $\epsilon_{NN}=0$), and
(c) log folding rate $k$ as a
function of the average non-native energy $\epsilon_{NN}$ (for $b=0$).
The parameters controlling the strength of the non-native energy, $b^2/2T$ and $\epsilon_{NN}$,
enter the free energy at the same footing in the theoretical model
(see equation~(\ref{eq:F})).
Results from simulations are in very good agreement with the theoretical prediction
(dashed black curve, in figures (b) and (c)) obtained when the value of $ \Delta A^{\prime\: \neq}$
--shown in (a)-- is used as an estimate of $\Delta A^{\neq} \equiv
\Del \Ast(Q^{\neq})$ entering equation~(\ref{dftf})
predicted by the theory. 
Values of $\ln k$ and $\Delta
F^{\dag}$ are normalized to the corresponding values for the unperturbed
case ($\ln k_0$ and $\Delta F_{0}^{\dag}$). For large non-native energy perturbations
($b > 1$, or $\epsilon_{NN}> 0.5$) both $\Delta F^{\dag}(T_f^0)$ and $\ln k$ rapidly 
decrease (see also figure~\ref{fig:cv}(b) and (c)).
The energy parameters $b$ and $\epsilon_{NN}$ are measured in
units of native energy per contact, $\epsilon$.  Barrier heights are
measured in units of the folding temperature for the unperturbed case
($\kboltz T_0$).

\vspace{0.6in}

\begin{figure*}[!h]
\centerline{\includegraphics[width = 0.9\linewidth]{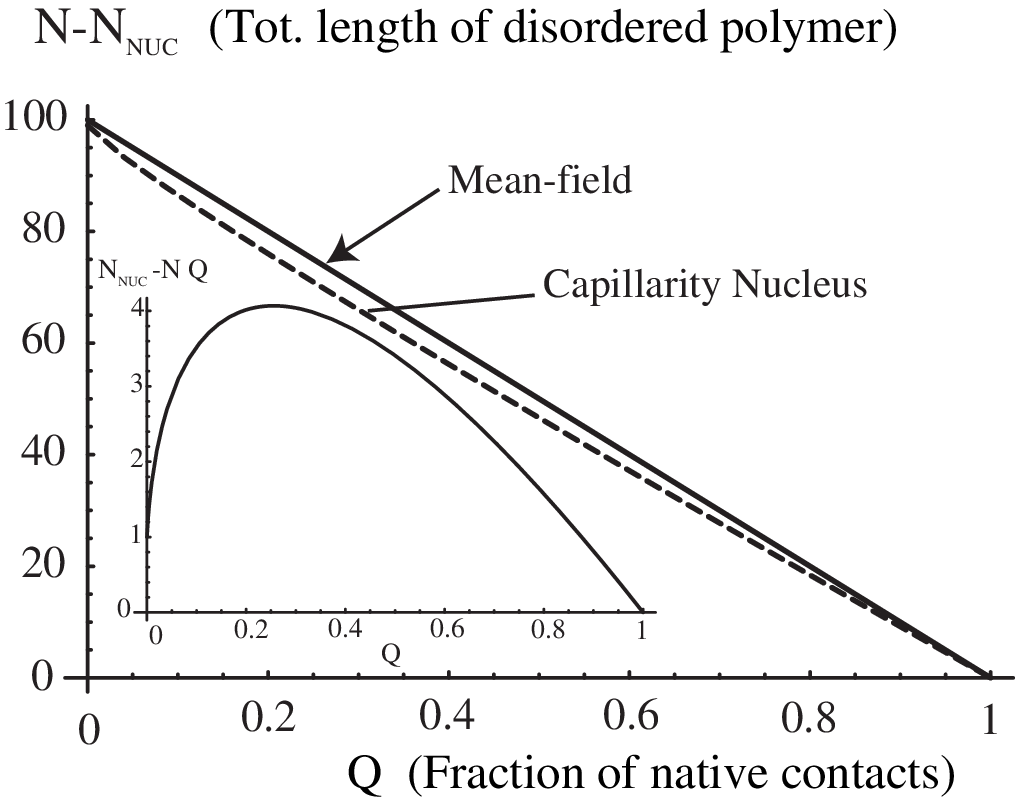}}
\caption{\sf\protect
}
\label{fignLQ}
\end{figure*}

FIGURE \ref{fignLQ}:
Plot of the amount of disordered polymer in the protein as a function
of $Q$, for a mean-field model which has the form $N (1-Q)$, and for a
capillarity model which has the form $N - \Nnuc$, where $\Nnuc$ is
given in equation~(\ref{cubic1}). The inset shows the difference on a
magnified scale. Here the chain length $N=100$, and the mean fraction
$\s$ of interactions lost on the surface of the capillarity nucleus
(see equation~(\ref{nuc1})) is taken numerically to be $1.0$.
For systems of size $100$ the deviation is only a few percent, the
relative deviation goes as $N^{-1/3}$.

\vspace{0.6in}

FIGURE \ref{fig_lj}:
Non-native interactions for increasing interaction strength (regulated by
the parameter $\eta$, see equation~\ref{expr_nn}), from highly repulsive to
highly attractive (thin curves). Curves for each value of
$\eta$ indicate a $1 \sigma$ width in the non-native potentials. The
unperturbed potential (short-range repulsive term in
equation~\ref{go_ham}) is plotted as a reference (thick curve).  Energies
are measured in units of native energy per contact, $\epsilon$.
\newpage

\centerline{FIGURES}
\vspace{0.5in}

\newpage

\newpage

\newpage

\newpage

\newpage

\newpage

\newpage

\newpage

\newpage

\newpage

\newpage

\end{document}